\documentclass[10pt]{iopart}
\usepackage[english]{babel}
\usepackage{csquotes}
\usepackage{iopams}
\usepackage{parskip}

\usepackage[square,sort&compress,comma,numbers]{natbib}
\usepackage[font=small,labelfont=bf]{caption} 
\usepackage{amsmath}
\usepackage{graphicx}
\usepackage{physics}
\usepackage{wrapfig}
\graphicspath{{images/}}
\usepackage{fancyhdr}
\usepackage{geometry}
\usepackage{comment}
\usepackage{amsbsy}
\usepackage{float}
\usepackage[colorlinks = true,
            linkcolor = blue,
            urlcolor  = blue,
            citecolor = blue,
            anchorcolor = blue]{hyperref}
\usepackage{array}
\usepackage{url}
\usepackage[noabbrev]{cleveref}
\newcolumntype{M}[1]{>{\centering\arraybackslash}m{#1}}
\allowdisplaybreaks

\begin{document}
\title[Paper]{High-resolution spectroscopy of a single nitrogen-vacancy defect at zero magnetic field}
\author{Shashank Kumar$^1$, Pralekh Dubey$^1$,
Sudhan Bhadade$^1$, Jemish Naliyapara$^1$, Jayita Saha$^1$, and Phani Peddibhotla$^1$ \footnote{Present address: Department of Physics, IISER Bhopal, India}}
\address{$^1$ Department of Physics,
Indian Institute of Science Education and Research Bhopal, Madhya Pradesh, India}
\ead{phani@iiserb.ac.in}

\begin{abstract}\\
We report a study of high-resolution microwave spectroscopy of nitrogen-vacancy centers in diamond crystals at and around zero magnetic field. We observe characteristic splitting and transition imbalance of the hyperfine transitions, which originate from level anti-crossings in the presence of a transverse effective field. We use pulsed electron spin resonance spectroscopy to measure the zero-field spectral features of single nitrogen-vacancy centers for clearly resolving such level anti-crossings. To quantitatively analyze the magnetic resonance behavior of the hyperfine spin transitions in the presence of the effective field, we present a theoretical model, which describes the transition strengths under the action of an arbitrarily polarized microwave magnetic field. Our results are of importance for the optimization of the experimental conditions for the polarization-selective microwave excitation of spin-1 systems in zero or weak magnetic fields. 
\end{abstract}
\noindent{\it Keywords\/}: nitrogen-vacancy center, zero-field electron spin resonance spectroscopy, quantum sensing

\setcounter{page}{1}

\section{Introduction} 
Nitrogen-vacancy (NV) center, a paramagnetic defect in diamond \cite{Schirhagl:2014}, is a highly versatile sensor with excellent sensitivity to magnetic, electric, and stress fields at room temperature ~\cite{Maze:2008,Dolde:2011,Barson:2017}. Owing to the well-established optical initialization and readout methods \cite{Harrison:2004,Gruber:1997}, the ground state spin of the NV center is amenable to sensing applications \cite{Doherty:2012}. The quantum sensing \cite{Degen:2017} properties of the NV centers are being exploited for applications in biology \cite{Davis:2018,Sage:2013}, geology \cite{Glenn:2017}, material science \cite{Lillie:2018,Simpson:2016}, condensed matter physics \cite{Casola:2018}, quantum computation, and information processing \cite{Childress:2013,Pezzagna:2021}. Typically, most of the sensing experiments involving NV centers require simultaneous application of an external bias magnetic field. This requirement is detrimental to applications including nano-NMR in weak magnetic fields \cite{Cerrillo:2021}, structural investigation of molecules \cite{Vetter:2022} and in the experiments performed in magnetically shielded environments \cite{Jarmola:2021}. However, at zero magnetic field, the interaction of the NV center with the intrinsic effective field \cite{Doherty:2012} comprising the local static electric field and the local deformation-induced strain field can dominate, resulting in the inaccurate estimation of the external magnetic field of interest. Therefore, it is crucial to thoroughly understand and characterize the local charge and stress environment of the NV center in diamonds, as these studies may pave the way for efficient engineering of the intrinsic properties of the NV defects in diamonds for zero-field sensing applications. Furthermore, careful understanding of the ground state Hamiltonian in the presence of spin-electric and spin-strain interactions \cite{Doherty:2012,Barson:2017} was instrumental for key advances, including the single NV-based electrometry \cite{Dolde:2011}, zero-field magnetometry \cite{Zheng:2019,Lenz:2021,Wang:2022}, nanospin-mechanical sensors \cite{Barson:2017}, optically enhanced electric field sensing \cite{Block:2021} and the experimental demonstration of holonomic quantum gates \cite{Nagata:2018}.\vspace{0.5cm}\\
In this work, we have studied the effects of the intrinsic effective field on the NV hyperfine level structure by performing high-resolution microwave spectroscopy on single NV centers in a polycrystalline diamond (PCD) sample. Previous studies involving the NV centers in PCD samples have demonstrated long ground state electron spin coherence times \cite{Jahnke:2012}, thereby showing great promise for applications in wide-field quantum sensing \cite{Trusheim:2016} and quantum information processing \cite{Bersin:2019}.  
By leveraging long-lived ground state spin coherence, we perform pulsed-optically detected magnetic resonance (p-ODMR) \cite{Dreau:2011} measurements to record the hyperfine resolved spin transitions of an NV center experiencing a large intrinsic effective field. The presence of effective field inside a diamond lattice perturbs the NV hyperfine level structure by mixing and splitting the $\ket{m_s = \pm1}$ states, which can be observed as a shifting and splitting of the hyperfine transitions involved. The shifting of the NV's overall spectrum from the central transition frequency of $\sim$ 2.87 GHz is attributed to the axial component of the effective field, whereas the splitting of the inner and/or outer transitions is attributed to the transverse component of the effective field. Recent studies to precisely characterize the effective fields surrounding NV spins \cite{Mittiga:2018,Kolbl:2019,Knauer:2020} revealed the mixing of the central hyperfine resonances in the zero-field spectrum for a single NV center. In contrast, the mixing of the outer hyperfine resonances was not observable because of the presence of low effective fields in their diamond crystals under study. In this work, our studies show that the presence of large intrinsic effective fields in PCD samples leads to the mixing of the NV's outer hyperfine transitions at zero magnetic field which, to the best of our knowledge, has not been observed experimentally until now. We also perform NV-based spectroscopic characterization of the local effective field environment in the PCD sample. Our studies conclude that the zero-field spectral features of the single NV centers in PCD sample are dominated by the effect of local strain fields rather than by the effect of local charges. Finally, we introduce a theoretical model of the magnetic dipole transitions that provides an improved understanding of the polarization response \cite{Kolbl:2019} of the hyperfine spin transitions at zero magnetic field.
\section{Theory}
\label{section:Theory}
\begin{figure}[h]
    \centering
    \includegraphics[scale=0.68]{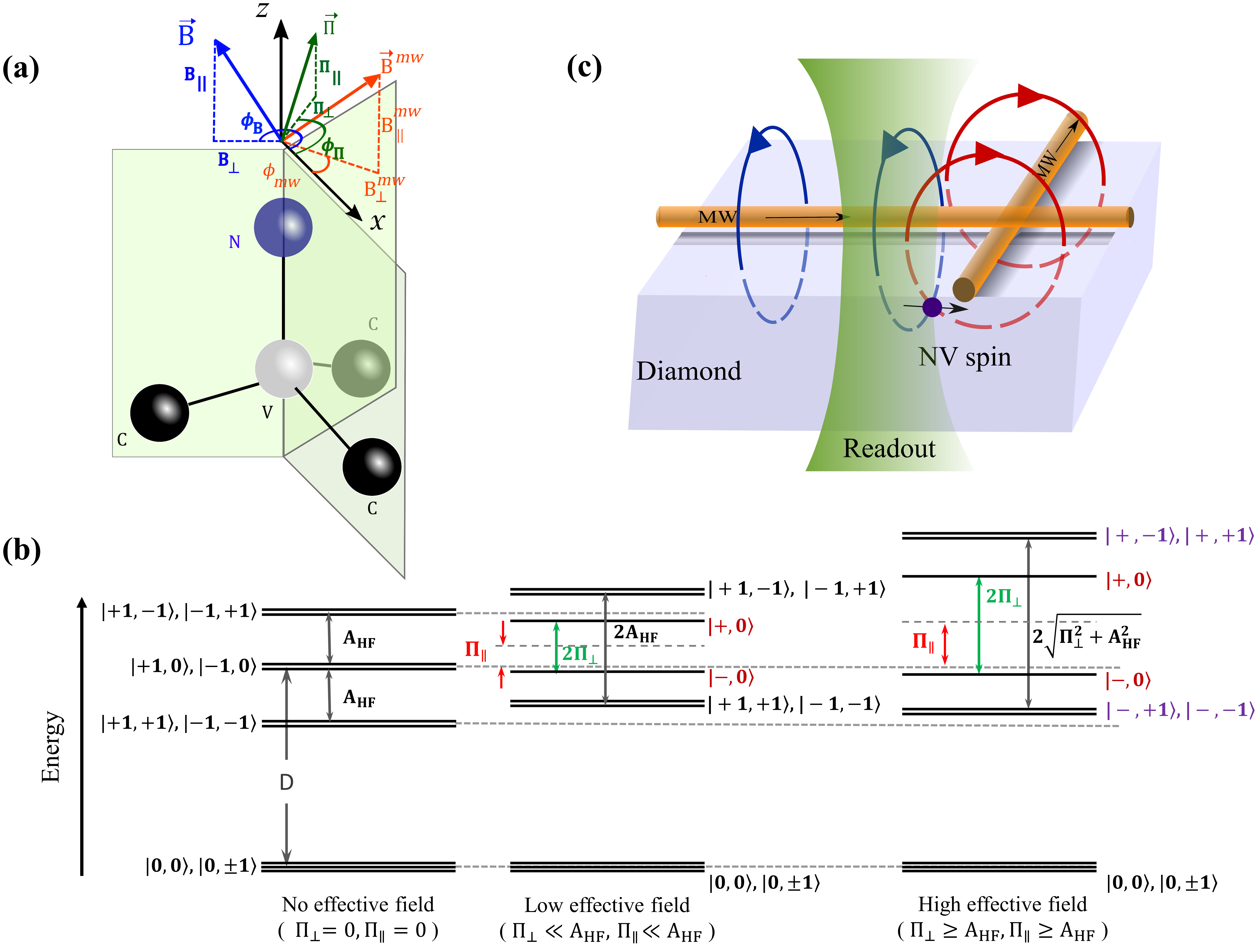}
    \caption{\textbf{(a)} Coordinate system of the negatively charged NV center in diamond. The blue sphere and the light gray sphere represents the NV center. The green vector represents the intrinsic effective field acting at the NV site. The red vector represents the microwave (MW) magnetic field used to manipulate the NV spin. The blue vector represents the external magnetic field. \textbf{(b)} Energy-level diagram for the NV center in diamond, showing the shifts and splittings of the energy levels of the electron-nuclear spin system due to the effective field interaction term in the Hamiltonian. The energy levels are plotted in three different parameter regimes: (i) $\Pi_{\perp}=\Pi_{\parallel}=0$, (ii) $A_{HF}\gg\Pi_{\perp},\Pi_{\parallel}$ and (iii) $A_{HF}\leq\Pi_{\perp},\Pi_{\parallel}$. \textbf{(c)} The MW antenna in a cross-wire configuration is used to drive the NV centers. This configuration enabled us to drive hyperfine resolved transitions of NV centers with two different orientations of the MW magnetic field relative to the NV quantization axis.}
    \label{fig:Energy_level_with_low_and_high_effective_field1}
\end{figure}

NV centers in diamond can exist in three charge states: neutral NV$^0$, positively charged NV$^+$, and negatively charged NV$^-$ states with different optical and spin properties \cite{Doherty:2012,Pfender:2017}. NV$^-$, hereafter referred to as NV for brevity, is a paramagnetic $S=1$ defect with two unpaired electrons in its ground and excited state. In the ground state, the  $m_s = 0$ and degenerate $m_s = \pm{1}$ spin sublevels are split by $D =2.87$ GHz, which is the fine structure splitting originating from the dipolar spin-spin interaction \cite{Rondin:2014}. A bias (or an external) magnetic field further splits the degenerate $m_s$ = $\pm{1}$ states due to the Zeeman interaction. The magnitude of the separation between the two states is proportional to the component of the applied magnetic field along the NV direction, and hence can be used for sensing small dc magnetic fields. However, when no external bias magnetic field is applied, the electric field originating from charge impurities and the strain field arising from local lattice deformations couples the otherwise degenerate $m_s$ = $\pm{1}$ states, leading to the mixing and splitting of the $m_s = \pm1$ states. These two local intrinsic fields are collectively named as the effective field ($\mathbf{\Pi}$) because they both arise from electronic interactions \cite{Mittiga:2018} that contribute to the variation of the spatial distribution of the electron density around the NV center. Moreover, the NV ground state spin is also coupled via the hyperfine interaction to the nuclear spin bath composed of the native nitrogen nuclear $^{14}N$ spin and naturally occurring nuclear $^{13}C$ spin impurities. In this study, we use an NV center coupled to its intrinsic $^{14}N$ spin to probe its local environment and demonstrate that the high-resolution, low-power optically detected electron spin resonance (ESR) spectroscopy can extract information about the spatial dependence of the axial and the transverse components of the effective field.\vspace{0.5 cm}\\ 
The ground state spin Hamiltonian $\mathcal{\hat{H}}$ of the NV center in the presence of the intrinsic effective field and the axial hyperfine field can be expressed as \cite{Sekiguchi:2016}
\begin{equation}
    \mathcal{\hat{H}} =D\hat S_z^2 + A_{HF} \hat S_z \hat I_z+\Pi_\parallel \hat S_z^2+\Pi_x \left(\hat S_x^2-\hat S_y^2\right)+\Pi_y\left(\hat S_x \hat S_y+\hat S_y\hat S_x\right)
    \label{eqn:Main_Hamiltonian_with_magnetic_field}
\end{equation}
where $D = 2.87$ GHz is the zero-field splitting parameter, $A_{HF} = -2.14$ MHz is the axial hyperfine coupling parameter, $\vb*{\hat{I}}=(\hat{I}_x,\hat{I}_y,\hat{I}_z)$ is the dimensionless nuclear spin-1 vector operator of the host $^{14}$N spin, $\vb*{\hat{S}}=(\hat{S}_x,\hat{S}_y,\hat{S}_z)$ is the dimensionless NV electronic spin-1 vector operator and $\vb*{\Pi} = (\Pi_x,\Pi_y,\Pi_\parallel)$ is the effective field vector expressed in the NV coordinate frame (xyz) where $z$ denotes the NV axis and $x$ lies in one of the symmetry planes as shown in figure \ref{fig:Energy_level_with_low_and_high_effective_field1}(a). The axial and non-axial components of the effective field are represented as $\Pi_\parallel=d_{\|}E_z+M_z $  and $\Pi_{x,y}=d_{\perp} E_{x,y}+M_{x,y}$ where $\vb*{E}=(E_x,E_y,E_z)$ is the electric field vector and $d_\parallel = 0.35 $ Hz cmV$^{-1}$ and $d_\perp = 17$ Hz cmV$^{-1}$ are the axial and transverse electric field susceptibilities, respectively \cite{Dolde:2011}. Here, $M_z$ and $M_{x,y}$ are the spin-strain interaction parameters that further depend on the spin-strain susceptibilities and, unlike electric field susceptibilities, are of comparable magnitude (refer to \cite{Barson:2017} for more details).\vspace{0.5cm}\\ 
Since the effective field introduces a coupling between the $m_s=\pm1$ spin states with the same hyperfine projection, the Hamiltonian in the equation (\ref{eqn:Main_Hamiltonian_with_magnetic_field}) can also be rewritten as
\begin{equation}
 \label{eqn:Transformed_main_Hamiltonian}
  \mathcal{\hat H}^{m_I} = (D +\Pi_\parallel)\hat{\mathbb{I}}+\Pi_x \hat{\sigma}_x^{m_I}+\Pi_y \hat{\sigma}_y^{m_I}+m_IA_{HF}\hat{\sigma}_z^{m_I}
\end{equation}
where $\hat{\mathbb{I}}$ denotes the identity operator and $\hat{\sigma}_x^{m_I}$, $\hat{\sigma}_y^{m_I}$ and $\hat{\sigma}_z^{m_I}$ denote the Pauli operators in the subspace $\{\ket{m_s, m_I}\}=\{\ket{+1, m_I};\ket{-1, m_I}\}$ \cite{Sekiguchi:2016}. Here, $m_s=0,\pm1$ and $m_I=0,\pm1$ are the eigenvalues of the $\hat{S}_z$ and $\hat{I}_z$ operators, respectively. Solving Hamiltonian (\ref{eqn:Transformed_main_Hamiltonian}) gives the eigenenergies of the hyperfine states in the presence of the intrinsic effective field as
\begin{subequations}
\begin{align}
 \label{eqn:Transformed_main_eigenenergies_1_1}
 {E}_{-,0}&= D + \Pi_{\parallel}-\Pi_{\perp}, \\
 {E}_{+,0}&= D + \Pi_{\parallel}+\Pi_{\perp}, 
\end{align}
\begin{align}
 \label{eqn:Transformed_main_eigenenergies_1_2}
 {E}_{-,+1}&= D + \Pi_{\parallel}-\sqrt{\Pi_{\perp}^2+\left(A_{HF}\right)^2}, \\
 {E}_{+,+1}&= D + \Pi_{\parallel}+\sqrt{\Pi_{\perp}^2+\left(A_{HF}\right)^2}, 
\end{align}
\begin{align}
 \label{eqn:Transformed_main_eigenenergies_1_3}
 {E}_{-,-1}&= D + \Pi_{\parallel}-\sqrt{\Pi_{\perp}^2+\left(A_{HF}\right)^2}, \\
 {E}_{+,-1}&= D + \Pi_{\parallel}+\sqrt{\Pi_{\perp}^2+\left(A_{HF}\right)^2}, 
\end{align}
\end{subequations}
where $\Pi_{\perp}=\sqrt{\Pi_x^2+\Pi_y^2}$ and $\Pi_{\parallel}=\Pi_\parallel$ are the transverse and parallel effective field amplitudes, respectively. The associated hyperfine eigenstates can be written in the uncoupled basis $\ket{m_s,m_I}=\ket{m_s}\otimes\ket{m_I}$ as follows:
\begin{subequations}
\label{eqn:All_main_eigenstates}
 \begin{align}
 \ket{-,0}&=\ket{-}_0\otimes\ket{0}=\frac{1}{\sqrt{2}}\left(\ket{+1}-e^{i\phi_\Pi}\ket{-1}\right)\otimes\ket{0}, \label{eqn:coupled_states_for_m_I=0_plus} \\
  \ket{+,0}&=\ket{+}_0\otimes\ket{0}=\frac{1}{\sqrt{2}}\left(\ket{+1}+e^{i\phi_\Pi}\ket{-1}\right)\otimes\ket{0} \label{eqn:coupled_states_for_m_I=0_minus},
\end{align}  
\begin{align}
\ket{-,+1}&=\ket{-}_{+1}\otimes\ket{+1}=\left(\sin{\frac{\theta_\Pi}{2}}\ket{+1}-e^{i\phi_\Pi} \cos{\frac{\theta_\Pi}{2}}\ket{-1}\right)\otimes\ket{+1} \label{eqn:Transformed_main_eigenstates_1},\\
\ket{+,+1}&=\ket{+}_{+1}\otimes\ket{+1}=\left(\cos{\frac{\theta_\Pi}{2}}\ket{+1}+e^{i\phi_\Pi} \sin{\frac{\theta_\Pi}{2}}\ket{-1}\right)\otimes\ket{+1} \label{eqn:Transformed_main_eigenstates_2},
\end{align}
\begin{align}
 \ket{-,-1}&=\ket{-}_{-1}\otimes\ket{-1}=\left(\cos{\frac{\theta_\Pi}{2}}\ket{+1}-e^{i\phi_\Pi} \sin{\frac{\theta_\Pi}{2}}\ket{-1}\right)\otimes\ket{-1} \label{eqn:Transformed_main_eigenstates_3}, \\
 \ket{+,-1}&=\ket{+}_{-1}\otimes\ket{-1}=\left(\sin{\frac{\theta_\Pi}{2}}\ket{+1}+e^{i\phi_\Pi} \cos{\frac{\theta_\Pi}{2}}\ket{-1}\right)\otimes\ket{-1}\label{eqn:Transformed_main_eigenstates_4},
\end{align}
\end{subequations}
where $\theta_\Pi$ and $\phi_\Pi$ are the angles which satisfy the relations
\begin{subequations}
    \begin{align}
     \cos{\theta_\Pi}&=\frac{-|A_{HF}|}{\sqrt{\Pi_{\perp}^2+\left(A_{HF}\right)^2}},\label{eqn:theta}\\
    \cos{\phi_\Pi} &= \frac{\Pi_x}{\Pi_{\perp}} \label{eqn:phi}.
    \end{align}
\end{subequations}
We note that the hyperfine eigenstates labeled with $\ket{0}_{m_I}=\ket{0,m_I}$ remain three-fold degenerate as their eigenenergies are not altered by the presence of the intrinsic effective fields. 
We can make the following conclusions regarding the nature of the mixed states and the theoretically expected ODMR spectra for samples experiencing different magnitudes of transverse effective field:
\begin{itemize}
\item Case 1: $\Pi_\perp \ll |A_{HF}|$\\
When the hyperfine coupling between the NV spin and the $^{14}N$ nucleus is much larger than the transverse component of the effective field, the states $\ket{m_s = \pm 1,m_I=0}$ couple to form two new eigenstates with a splitting of $2\Pi_\perp$. In contrast, the mixing of the states  $\ket{-1,+1}$ \& $\ket{+1,+1}$ and $\ket{-1,-1}$ \& $\ket{+1,-1}$ due to the effective field is suppressed by the hyperfine coupling parameter $A_{HF}$. Hence, the outer transition frequencies are not affected by the effective field, i.e. the hyperfine projections with the same $m_I$ are simply split by $2|A_{HF}|$ as it is for the case of the Hamiltonian without the effective field (refer \cref{fig:Energy_level_with_low_and_high_effective_field1}(b)).
\item Case 2: $\Pi_\perp \gtrsim |A_{HF}|$\\
If the transverse component of the effective field is comparable to the hyperfine splitting, the states  $\ket{-1,+1}$ \& $\ket{+1,+1}$ and $\ket{-1,-1}$ \& $\ket{+1,-1}$ also couple to form two new eigenstates with a splitting of $2\sqrt{\Pi_\perp^2+A_{HF}^2}$. Relations defined in {\crefrange{eqn:Transformed_main_eigenstates_1}{eqn:Transformed_main_eigenstates_4}} show that the mixing between the states depends not only on the azimuthal angle $\phi_\Pi$ of the transverse effective field, but also on the angle $\theta_\Pi$ which quantifies the relative strengths of hyperfine interaction and transverse effective field.
\end{itemize}
\section{Experiments}
\label{section:Experiments}
\subsection{\textbf{Description of the experimental setup}}
\label{subsection:Description_of_experimental_setup}
We used a home-built confocal microscopy setup to optically address the single NV centers in diamond at room temperature. A 532 nm green laser was used to excite the diamond through a high numerical-aperture (NA) oil immersion microscope objective (Olympus, 100$\times$ NA = 1.40). The red fluorescence emitted by the single NV centers is collected by the same objective lens and then focused through a pinhole of 50 microns diameter before being detected by a single photon counting module (SPCM). A signal generator (Keysight N5171B) is used as a source for generating microwave (MW) fields and the MW fields are amplified with the help of an MW amplifier (Mini-Circuits, ZHL-16W-43-S+). The MW fields were applied to the NV centers through two 25$\mu$m thick copper wires arranged in a cross-configuration as shown in figure \ref{fig:Energy_level_with_low_and_high_effective_field1}(c), bridged across the diamond surface.\vspace{0.5cm}\\
In order to perform experiments in the zero magnetic field regime, the cancellation of the stray magnetic field acting along the NV defect axis was performed with the help of a neodymium (NdFeB) permanent magnet placed on a $xyz$-stage close to the diamond sample. To understand the influence of the magnetic field on the NV hyperfine level structure, it is illustrative to resolve the weak field ($B <$ 1 G) from the magnet into components parallel and perpendicular to the NV symmetry axis. As we know, the NV spin is much more susceptible to the axial magnetic fields than the transverse magnetic fields. Hence, the parallel component is strong enough to couple to the NV spin, while the perpendicular component is not strong enough to yield a significant second order shift of the NV spin energies \cite{Thiel:2016}. Hence, the permanent magnet can be aligned to a desired orientation such that it's magnetic field nearly cancels the stray magnetic field acting along the NV axis. At the same time, the transition energies of individual NV centers will be minimally affected by the magnetic fields transverse to the NV axis. For every NV center investigated in this work, we applied this method to obtain a series of hyperfine spectra and the spectra were fitted to estimate the value of the residual axial magnetic field.  
\subsection{\textbf{High-resolution spectroscopy}}
\label{subsection:High-resolution_spectroscopy}
We investigated single NV spins in a type-IIa polycrystalline diamond sample and type-Ib single crystal diamond sample to compare our experimental measurements with theoretical calculations. The discussed features regarding the coupling of the hyperfine states due to the effective field are present in the magnetic spectra of the single NV spins in these samples. We used the ODMR technique  to study the ground-state electron spin transitions and to resolve the anti-crossings between the hyperfine levels, which arise from the interaction of the electron spin with the effective fields that are intrinsic to the diamond lattice. High resolution ODMR spectra \cite{Dreau:2011} were obtained by sweeping the linearly polarized microwave frequency across the hyperfine spin sublevel transitions of the NV center. ODMR spectra were fitted with an appropriate number of Gaussian functions to obtain the ESR transition frequencies. Subsequently, we used the extracted frequency values to determine the Hamiltonian parameters $\Pi_\perp$ and $\Pi_\parallel$ from the eigenvalue equations (\ref{eqn:Transformed_main_eigenenergies_1_2}) by applying the least square method.
\subsubsection{\textbf{Zero-field spectroscopy}}
\label{subsubsection:At_zero_magnetic_field} \leavevmode  \vspace{0.0001cm}

\begin{figure}[ht]
    \centering
    \includegraphics[scale=0.69]{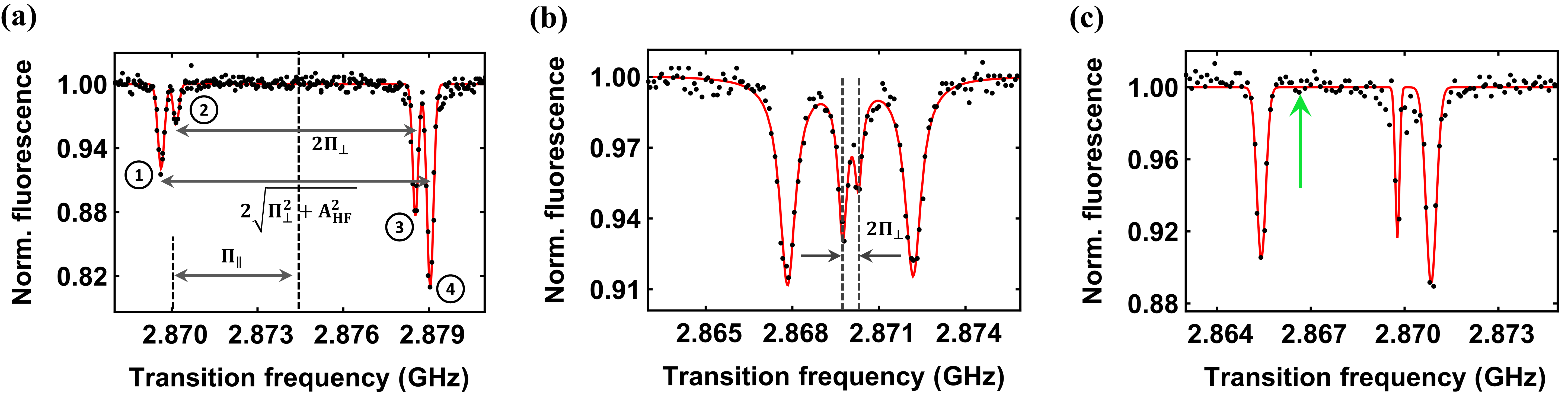}
    \caption{\textbf{(a)} High-resolution ODMR spectrum of a single NV center experiencing a large strain in the PCD sample. The spectrum shows a large shifting and splitting of both the inner and the outer transitions. The transverse effective field induces a splitting of $2\Pi_\perp \approx 8.38 \;MHz$, while the axial effective field induces a common-mode shift of $\Pi_\parallel \approx 4.32\; MHz$ in the hyperfine energy levels. The observed values of the imbalance for the outer and inner hyperfine resonances are $\mathcal{I}_{outer}\approx$ 37.4$\%$ and $\mathcal{I}_{inner}\approx$ 50.8$\%$ respectively. \textbf{(b)} High-resolution ODMR spectrum of a single NV center in a type Ib diamond sample. The central line is shifted by $\Pi_\parallel \approx 50\; kHz$ and is further split by $2\Pi_\perp \approx 500 \;kHz$ with a transition imbalance of $\mathcal{I}_{inner}\approx$ -19.8$\%$. For type Ib samples, the presence of the electric fields originating from charge impurities leads to a splitting and imbalance of the inner hyperfine resonances. For the outer transitions, the additional splitting (2$\sqrt{\Pi_\perp^2+A_{HF}^2}-2A_{HF}$) and transition imbalance ($\propto\sin{\theta_\Pi}$) induced by the electric field is hardly discernible because $A_{HF}\approx10\Pi_{\perp}$. The negligible mixing of the outer states also means that the outer transitions exhibit a purely circularly polarized response. \textbf{(c)} Dark-state ODMR spectroscopy of a single NV center in a PCD sample. The green arrow shows the expected position of the perfect dark state, which does not interact with the microwave magnetic field. Here, the overall spectrum experiences a common-mode shift of $\Pi_\parallel \approx -1.85 \; MHz$ and the central transition is expected to be split by $ 2\Pi_\perp \approx 3.3\;MHz$.}
    \label{fig:PCD+HPHT+3dips+5dips}
\end{figure}

Figure \ref{fig:PCD+HPHT+3dips+5dips}(a) shows our high-resolution spectroscopy study of the effective field for a single NV center at zero magnetic field in a polycrystalline diamond sample, in which the spectrum is expected to be dominated by the effect of local strain \cite{Trusheim:2016}. Four distinct features are seen in the ODMR spectrum corresponding to two two-fold degenerate outer hyperfine transitions and two inner non-degenerate hyperfine transitions. As discussed in section \ref{section:Theory}, the two central resonances correspond to $m_I=0$ states with a separation of 2$\Pi_\perp$ and the outer resonances correspond to $m_I=\pm1$ states with a separation 2$\sqrt{\Pi_\perp^2+A_{HF}^2}$ owing to the presence of a large transverse effective field in the diamond sample. Moreover, the splittings of the hyperfine states are accompanied by a common-mode shift of all nuclear spin projections due to the presence of an axial effective field $\Pi_\parallel$. The spectra obtained are in very good agreement with our theoretical calculations, where the outer transitions are also mixed due to the presence of a large effective field (i.e. $\Pi_ \perp>|A_{HF}|$). The transverse effective field and the axial effective field extracted from the data shown in figure \ref{fig:PCD+HPHT+3dips+5dips}(a) are $\Pi_\perp = 4.20$ MHz and $\Pi_\parallel = 4.32$ MHz, respectively. \vspace{0.5cm}\\
We also performed similar measurements on a single NV defect in an untreated type-Ib diamond, and the recorded ODMR spectra are shown in figure \ref{fig:PCD+HPHT+3dips+5dips}(b). By analyzing the experimental data, we find $\Pi_\perp = 500$ kHz and $\Pi_\parallel = 50$ kHz. Hence, we note that that only the pair of inner transitions is significantly affected by the interaction of the NV spin with the effective field, as they are more susceptible to transverse effective fields than the outer transitions. Unlike the case of a strained type-IIa polycrystalline diamond sample, the observed zero-field spectral features of a single NV center in a type Ib sample are attributed to the local electric field \cite{Mittiga:2018} generated by the charge environment of the diamond lattice.\vspace{0.5cm}\\ 
\noindent\vspace{0.5cm}\\ 
The transition strengths observed in the high-resolution ODMR spectra depend on the overlap between the polarization of the applied microwave magnetic field and the magnetic dipole moment of the hyperfine spin transitions (see \ref{subsection:Magnetic_Dipole_Transition_Strengths}). The normalized transition strengths of the inner and outer transitions for a linearly polarized microwave field are given by: 
\begin{subequations}
\label{eqn:Transition_amplitudes_linear}
\begin{align}
  W_{\pm,0}&= \frac{1}{2} \left[1\pm \cos{(2\phi_{mw}-\phi_\Pi})\right],\\
  W_{-,\pm 1}&= \frac{1}{2} \left[1- \sin{\theta_{\Pi}}\;\cos{(2\phi_{mw}-\phi_\Pi})\right],\\
  W_{+,\pm 1}&= \frac{1}{2} \left[1+ \sin{\theta_{\Pi}}\;\cos{(2\phi_{mw}-\phi_\Pi})\right], 
\end{align}
\end{subequations}
where $W_{-,0}$ and $W_{+,0}$ are the normalized transition strengths corresponding to the transitions $\ket{0,0}\rightarrow \ket{-,0}$ and, $\ket{0,0}\rightarrow \ket{+,0}$ respectively. Similarly, $W_{-,+1}$, $W_{+,+1}$, $W_{-,-1}$  and $W_{+,-1}$ are the normalized transition strengths corresponding to the transitions $\ket{0,+1}\rightarrow \ket{-,+1}$, $\ket{0,+1}\rightarrow \ket{+,+1}$, $\ket{0,-1}\rightarrow \ket{-,-1}$ and $\ket{0,-1}\rightarrow \ket{+,-1}$ respectively.
Here, $\phi_{mw}$ and $\phi_{\Pi}$ are the azimuthal angles in the $xy$-plane (see figure \ref{fig:Energy_level_with_low_and_high_effective_field1}(a)), and $\theta_\Pi$ satisfies the relation (\ref{eqn:theta}). Since the outer transitions are degenerate, the normalized transition strengths for the outer transitions add together, resulting in twice the ODMR fluorescence contrast compared to a single transition. The transition imbalances \cite{Mittiga:2018,Kolbl:2019} of the inner and the outer hyperfine resonances are given by:
\begin{subequations}
\label{eqn:Imbalance}
\begin{align}
    \mathcal{I}_{inner} &= \cos(2\phi_{mw}-\phi_\Pi),\\
    \mathcal{I}_{outer} &=\sin\theta_{\Pi} \;\cos(2\phi_{mw}-\phi_\Pi)
\end{align}
\end{subequations}
This explains the measured difference in the relative strength of the two inner transitions and the two outer transitions, which is clearly seen in the data shown in the figure \ref{fig:PCD+HPHT+3dips+5dips}(a), where all the four resolvable transitions show different transition strengths. In contrast, for the data presented in figure \ref{fig:PCD+HPHT+3dips+5dips}(b), the transition imbalance is observed only for the central transitions and not for the two outer transitions.\vspace{0.5cm}\\ 
According to the equations (\ref{eqn:Transition_amplitudes_linear}), the transition strengths of the inner and outer hyperfine spin transitions under driving by a linearly polarized magnetic field oscillate as a function of the relative azimuthal angles $\phi_{\Pi}$ and $\phi_{mw}$. Hence, by characterizing the effective fields of several single NV centers, we found an NV center for which the normalized coupling strength between the $\ket{0,0}$ and $\ket{-,0}$ states is $\sim$1 resulting in the formation of a perfectly bright state and a perfectly dark state. The ODMR spectroscopy data shown in the figure \ref{fig:PCD+HPHT+3dips+5dips}(c) \cite{Footnote1} corresponds to one such NV center for which the linearly polarized microwave field does not interact with the transition magnetic dipole moment between the states $\ket{0,0}$ and $\ket{+,0}$. Hence, the effective field mixes the original $\ket{\pm1,0}$ states into perfectly bright and dark states $\ket{B}=\ket{-,0}$ and $\ket{D}=\ket{+,0}$ states respectively and the transitions $\ket{0,0}\rightarrow \ket{B}$ and $\ket{0,0}\rightarrow \ket{D}$ show the familiar linearly polarized response. In contrast, for the outer transitions under the same experimental conditions, the transition strengths are observed to be non-zero, which is ultimately linked to the elliptically polarized response \cite{Kolbl:2019} of the transitions  $\ket{0,\pm1}\rightarrow \ket{\pm,\pm1}$. We note that the fully bright and dark states can still be observed for the outer transitions via excitation with an elliptically polarized microwave field of suitably chosen ellipticity (see \cref{fig:Transition_strength_for_different_theta_and_epsilon} of \ref{subsection:Magnetic_Dipole_Transition_Strengths}).
\begin{figure}[ht]
    \centering
    \includegraphics[scale=0.75]{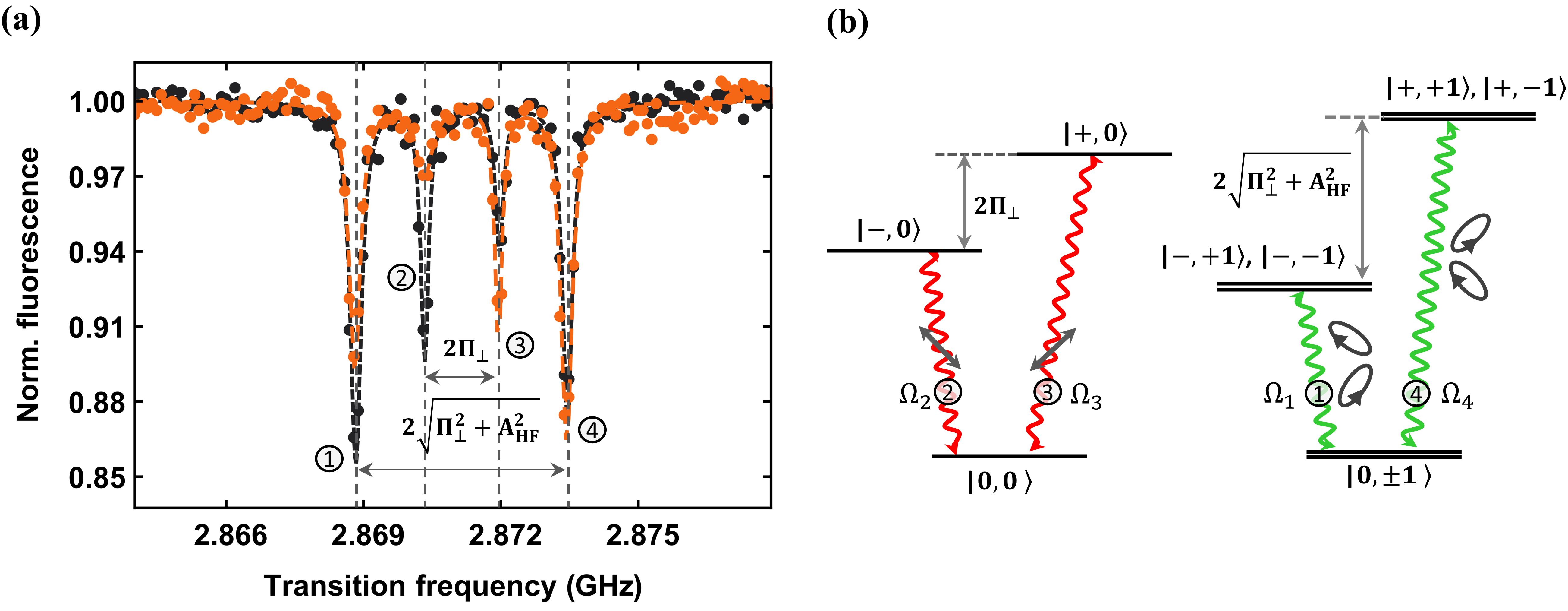}
    \caption{\textbf{(a)} ODMR spectra showing hyperfine resolved spin transitions for two possible azimuthal angles of the MW polarization interacting with the NV spin. \textbf{(b)} Polarization response of the hyperfine spin transitions. The inner transitions 2 \& 3 have a linearly polarized response, implying that a linearly polarized MW field is sufficient to achieve full polarization control of the coupling strength between the states $\ket{0,0}$ and $\ket{\pm,0}$. In contrast, the outer transitions 1 \& 4 have an elliptically polarized response, implying that an elliptically polarized MW field is necessary to achieve full control over the relative coupling strengths between the states $\ket{0,\pm1}$ and $\ket{\pm,\pm1}$ (see \ref{subsection:Magnetic_Dipole_Transition_Strengths}). Note that all the transitions follow the selection rule $\Delta m_I= 0$.}
    \label{fig:polarisation_port1_and_2_and_polarization_response}
\end{figure}
\subsubsection{\textbf{Zero-field spectroscopy for two different MW polarization angles}}
\label{subsubsection:At_zero_field_with_changing_mw_polarisation_angle} \leavevmode  \vspace{0.0001cm}

We also performed high-resolution spectroscopy of a single NV center for two different MW polarization angles $\phi_{mw}$ to study the influence of the relative orientation of the externally applied microwave field to the transition magnetic dipole moment. This was done by transmitting the microwaves through only one of the two crossed wires at a time as shown in the figure \ref{fig:Energy_level_with_low_and_high_effective_field1}(c). The two wires placed perpendicular to each other allowed us to probe the response of the NV spin transitions for two different polarizations of the MW drive. Since the transition imbalances for $m_I=0,\pm1$ depends on the MW angle, as is evident from the equations (\ref{eqn:Imbalance}), the observed ODMR response is different for each MW polarization under study. Figure \ref{fig:polarisation_port1_and_2_and_polarization_response}(a) show the experimental data on the same NV center, where the application of MWs with two different azimuthal angles reversed the sign of the transition imbalances for all the nuclear spin projections $m_I=0,\pm1$. However, the polarization angle of the MW magnetic field has no effect on the NV's ODMR transition frequencies as expected. Figure \ref{fig:Transition_strength_for_different_theta_and_epsilon} of \ref{subsection:Magnetic_Dipole_Transition_Strengths} depicts the transition strengths of the inner and outer transitions as a function of the MW angle.

\subsection{\textbf{Relation between Rabi frequencies and transition strengths}}
\begin{figure}[htbp]
    \centering
    \includegraphics[scale=0.85]{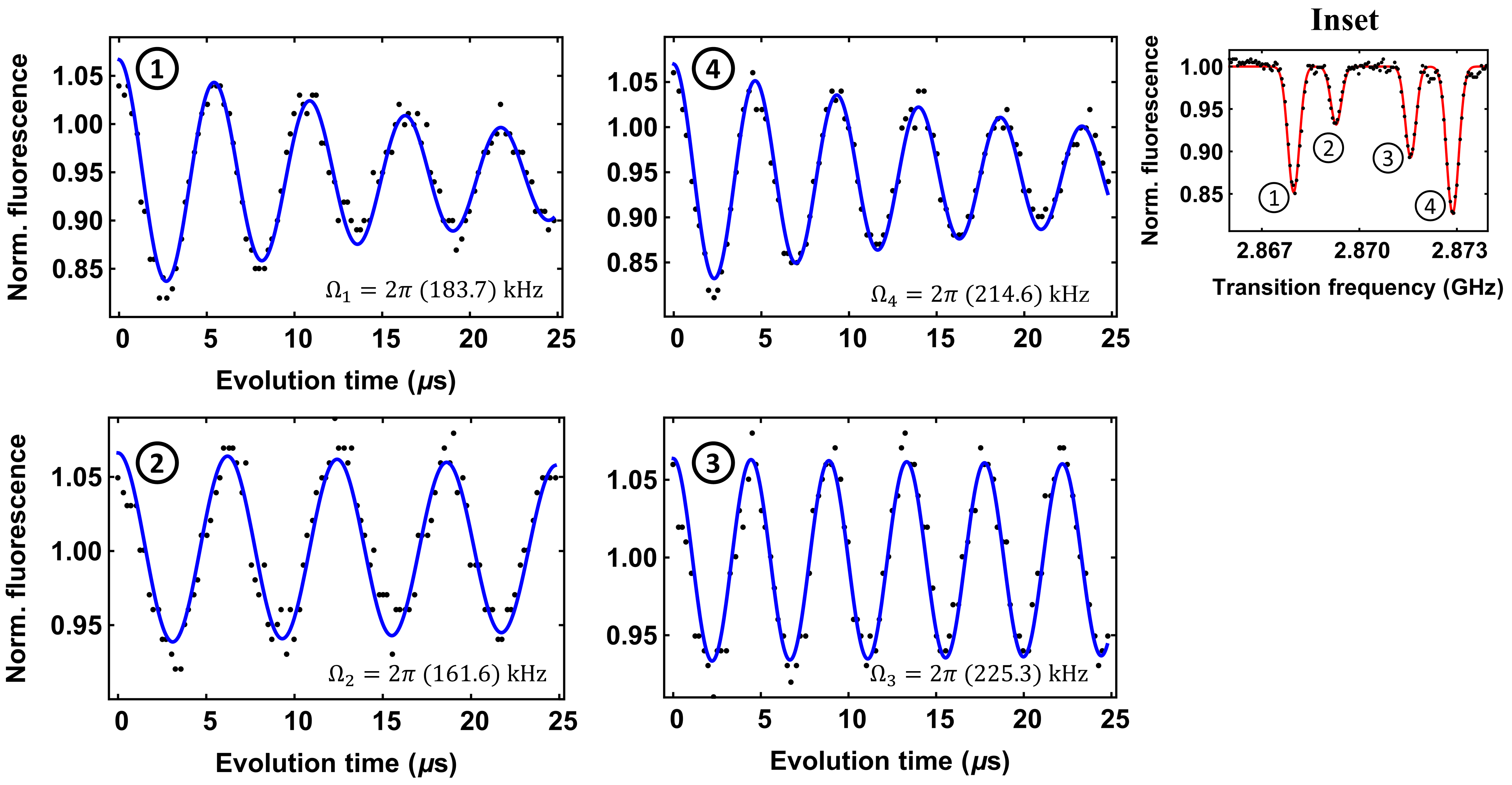}
    \caption{Rabi oscillation experiments performed to determine the transition strengths for each of the four hyperfine resonances of the pulsed ODMR spectrum presented in the inset of the figure. The Rabi frequencies of the four different transitions estimated from the experimental data satisfy the relation $\Omega_1^2 +\Omega_4^2 = \Omega_2^2+\Omega_3^2$ to a good approximation, where the Rabi frequencies $\Omega_1$ and $\Omega_4$ correspond to the two outer transitions and the Rabi frequencies $\Omega_2$ and $\Omega_3$ correspond to the two inner transitions.}
    \label{fig:rabi_osci}
\end{figure}

We performed Rabi oscillation experiments at zero magnetic field to determine the transition dipole moment for each of the four hyperfine transitions under the same experimental conditions, i.e. with the same MW and laser power, to compare the driving strengths of the involved ODMR transitions. By selectively driving each of the hyperfine resolved transitions with low microwave power, we measured the Rabi oscillations of all the four resonances of the ODMR spectrum shown in the inset of the figure \ref{fig:rabi_osci}. The relation between the Rabi frequencies of the four different hyperfine transitions is given by the equation (refer to \ref{subsection:Magnetic_Dipole_Transition_Strengths} for more information)
\begin{equation}
    \label{eqn:Relation_between_Rabi_frequencies}
    \Omega_1^2+\Omega_4^2=\Omega_2^2+\Omega_3^2
\end{equation}
where $\Omega_1$, $\Omega_2$, $\Omega_3$, and $\Omega_4$ are the respective Rabi frequencies of the hyperfine resonances 1, 2, 3 and 4 as marked in the inset of the figure \ref{fig:rabi_osci}. The experimental values for $\Omega_1$, $\Omega_2$, $\Omega_3$, and $\Omega_4$ satisfy the theoretical relation (\ref{eqn:Relation_between_Rabi_frequencies}) within 4$\%$ error. Note that although the degeneracy of the outer transitions results in a summation of the transition strengths for $m_I=\pm1$ hyperfine projections as observed in the  p-ODMR spectra, the measured Rabi frequencies $\Omega_1$ and $\Omega_4$ of the hyperfine spin transitions should not be affected by the degeneracy.
\noindent


\begin{figure}[h]
    \centering
    \includegraphics[scale=0.75]{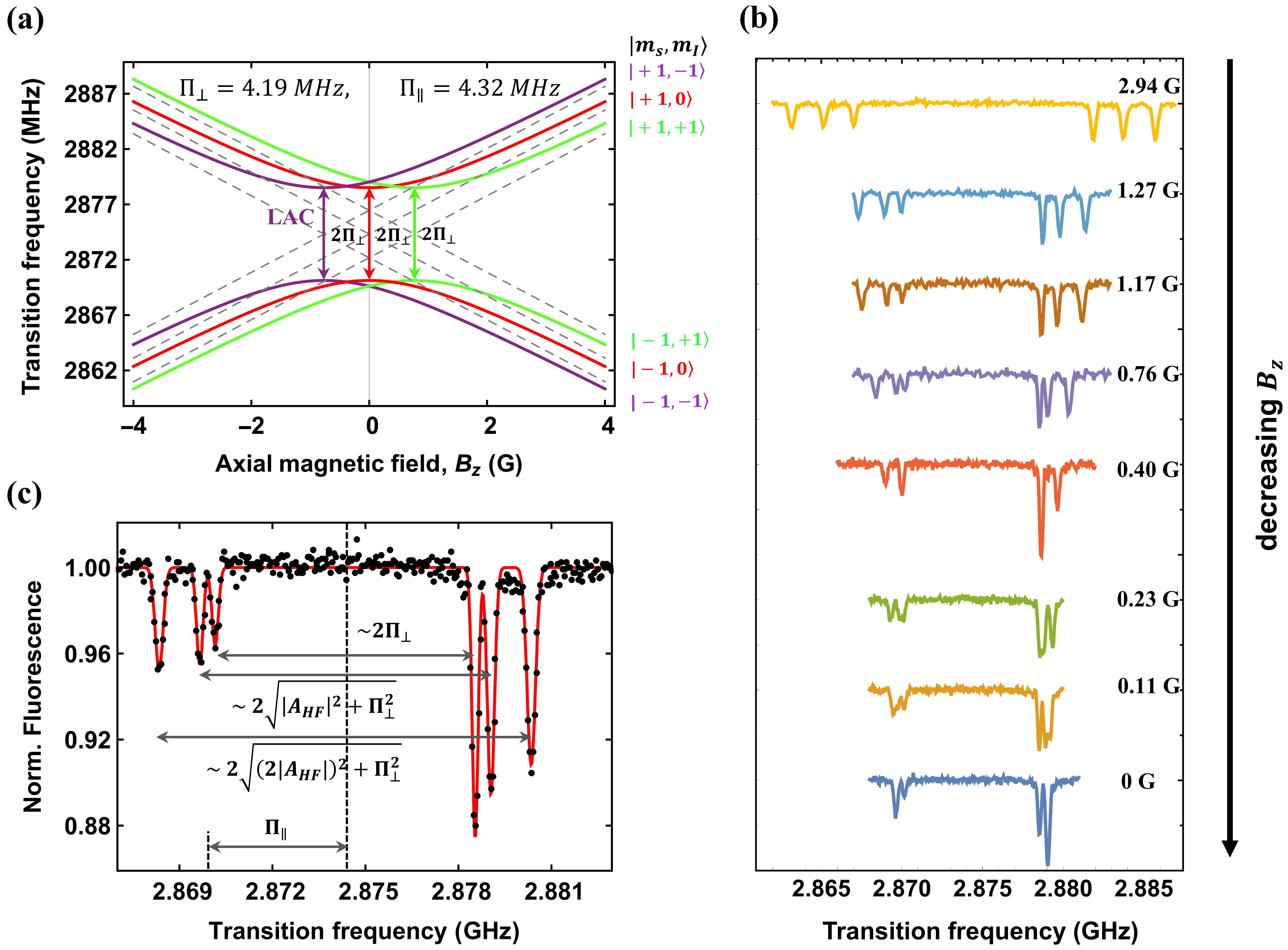}
    \caption{\textbf{(a)} Simulation plots of six possible transition frequencies from $\ket{0}_{m_I}$ to $\ket{\pm}_{m_I}$ as a function of axial magnetic field for an NV center, with the effective field parameters being $\Pi_\perp = 4.19 $ MHz and $\Pi_\parallel = 4.32 $ MHz. The gray dashed lines correspond to the transition frequencies in the absence of a transverse effective field, i.e. at $\Pi_\perp=0$ MHz and $\Pi_\parallel = 4.32 $ MHz. The colored arrows show the values of the axial magnetic field $B_z=m_I|A_{HF}|/\gamma$ where $m_I\in\{-1,0,+1\}$ at which the level anti-crossings (LACs) occur between the states $\ket{m_s=-1, m_I=0,\pm1}$ and $\ket{m_s=+1, m_I=0,\pm1}$ due to the interaction with the transverse effective field. \textbf{(b)} Experimental ODMR spectra for selected values of axial magnetic field $B_z$. \textbf{(c)} High-resolution spectrum of a single NV center in a PCD sample near the LAC $\gamma B_z \approx |A_{HF}|$, showing the mixing and splitting of the $m_I = +1$ transitions induced by the interaction with the effective field.}
    \label{fig:Simulations+ODMR_with_changing_magnetic_field}
\end{figure}
\subsection{\textbf{Spectroscopy around zero magnetic field}}
\label{subsubsection:At_finite_magnetic_field}
The Hamiltonian in equation (\ref{eqn:Transformed_main_Hamiltonian}) gets modified in the presence of an axial magnetic field and its matrix representation becomes:
\begin{equation}
    \renewcommand{\arraystretch}{1.3}
    \mathcal{{H}}^{m_I} = 
   \begin{bmatrix}
        (D +\Pi_\parallel)+m_I A_{HF}+\gamma B_z & \Pi_\perp e^{-i\phi_{\Pi}}\\
        \Pi_\perp e^{i\phi_{\Pi}} &(D +\Pi_\parallel) -m_I A_{HF}-\gamma B_z
    \end{bmatrix} 
\end{equation}
When there is no perturbation (i.e. $\Pi_\perp = 0$), the off-diagonal matrix elements of the above Hamiltonian vanish and the eigenvalues of the resulting Hamiltonian will be $E_{\pm,m_I} = (D +\Pi_\parallel)\pm (m_I A_{HF}+\gamma B_z)$, where $m_I\in\{-1,0,+1\}$. This results in a linear Zeeman splitting between the states $\ket{+ 1, m_I}$ and $\ket{-1, m_I}$. The simulated transition frequencies for this case are shown by gray dashed lines in figure \ref{fig:Simulations+ODMR_with_changing_magnetic_field}(a), where the levels cross at axial magnetic fields given by $B_z=m_I|A_{HF}|/\gamma$ in the absence of perturbation. However, when the Hamiltonian possesses non-diagonal matrix elements, i.e. for the case $\Pi_\perp \neq 0$, the energies of the two perturbed levels are given by $E_{\pm,m_I} = (D +\Pi_\parallel)\pm \sqrt{(m_I A_{HF}+\gamma B_z)^2 + \Pi_\perp^2}$. In other words, the perturbation mixes and splits the states $\ket{\pm1,m_I}$, and the level repulsion or anti-crossing is observed with the minimum separation between the perturbed energy levels being $2\Pi_{\perp}$. Colored plots in figure \ref{fig:Simulations+ODMR_with_changing_magnetic_field}(a) shows the simulated transition frequencies in the presence of the coupling $2\Pi_{\perp}$. The colored arrows illustrate the values of the axial magnetic field $B_z=m_I|A_{HF}|/\gamma$ at which the level anti-crossings (LACs) are observed between the states $\ket{\pm1,m_I}$. \vspace{0.5 cm}\\
In the following, we discuss the spectroscopy experiments performed around zero magnetic field. The p-ODMR spectra were recorded as a function of the decreasing axial magnetic field on the same NV center, whose zero field spectra was presented in figure \ref{fig:PCD+HPHT+3dips+5dips}(a). This procedure not only served the purpose of zeroing the axial component of the magnetic field, but also enabled us to perform high-resolution ODMR measurements in the vicinity of the transition crossings and anti-crossings. Figure \ref{fig:Simulations+ODMR_with_changing_magnetic_field}(b) shows the experimental p-ODMR spectra acquired for various values of the axial magnetic field $B_z$. The anti-crossings at $B_z = 0$ and $B_z = |A_{HF}|/\gamma$ can be visualized in the spectra shown in blue (first from bottom) and purple (fifth from bottom) respectively. The p-ODMR spectrum acquired at $ B_z \approx |A_{HF}|/\gamma$ is presented in more detail in figure \ref{fig:Simulations+ODMR_with_changing_magnetic_field}(c) and the spectrum shows a clear mixing and splitting of the states $\ket{\pm1,+1}$.

\subsection{\textbf{Characterization of the polycrystalline diamond}}
\label{section:PCD characterization}
\begin{figure}[h]
    \centering
    \includegraphics[scale=0.76]{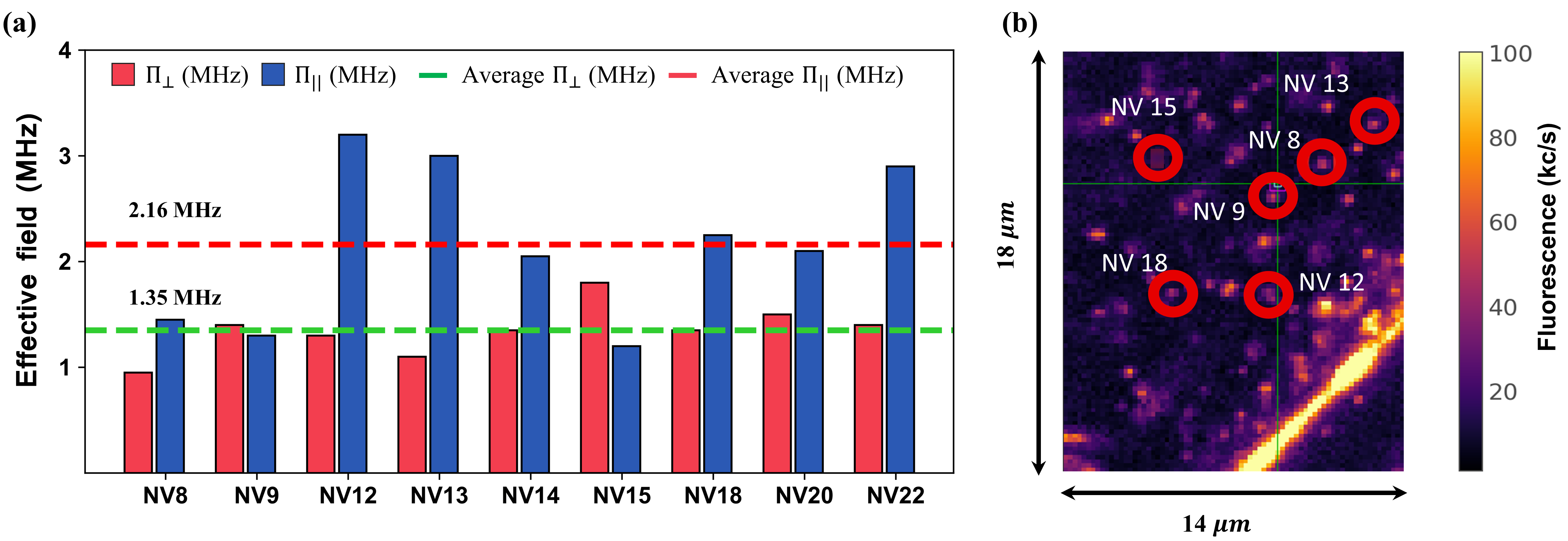}
    \caption{\textbf{(a)} Statistical investigation of the intrinsic effective fields of a selection of single NVs near the grain boundary in a PCD sample. The statistical average value of $\Pi_\perp$  is 1.35 MHz and that of $\Pi_\parallel$  is 2.16 MHz, which are of the same order of magnitude. These observations indicate that the strain field is the dominant source of the effective field near the grain boundary. \textbf{(b)}  14 $\mu$m $\times$ 18 $\mu$m confocal scan of the PCD sample showing some of the examined NVs close to the grain boundary.}
    \label{fig:stat}
\end{figure}

\begin{figure}[ht]
    \centering
    \includegraphics[scale=0.98]{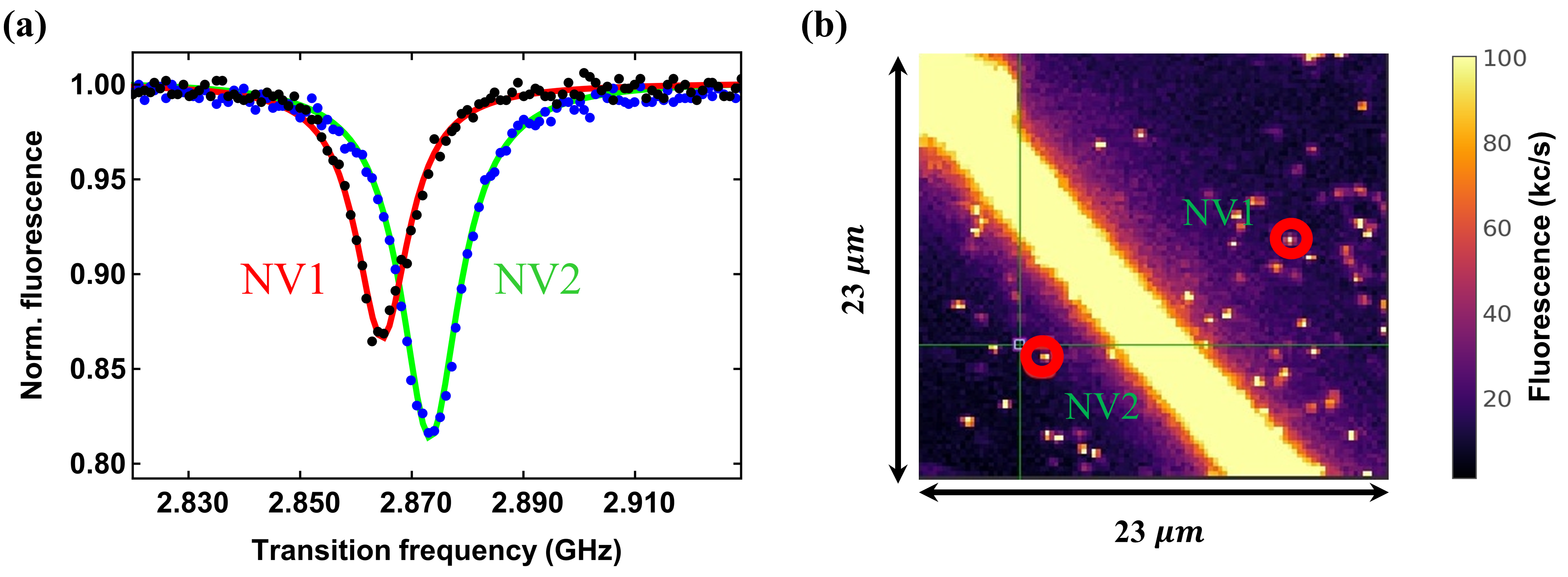}
    \caption{\textbf{(a)} CW ODMR spectra of NV1 and NV2, showing the negative and positive shifting of the overall spectrum by $\approx$ 5.31 MHz and $\approx$ 3.29 MHz respectively from 2.87 GHz. \textbf{(b)} 23 $\mu$m $\times$ 23 $\mu$m confocal scan of the PCD sample showing the highly fluorescent grain boundary and single NV centers on either side of it. The two NVs are marked in red circles.}
    \label{fig:Positive_&_negative_shift_CW_ODMR_and_Scan}
\end{figure}

Having established our high-resolution spectroscopy technique to measure both the parallel and perpendicular components of the effective field of a single NV spin, we now apply this method to characterize the effective field environment surrounding NV spins in different regions of the diamond sample. The observations are as follows:
\begin{itemize}
    \item The data of figure (\ref{fig:stat}) shows the effective field environment experienced by a selection of single NV centers over a distance of $10\mu m$ from the grain boundary. The splitting $2\Pi_\perp$ and the shifting $\Pi_\parallel$ for the individual NV centers are of comparable magnitude. Hence, we conclude that the effective field in a PCD sample is dominated by the strain field rather than by the electric field \cite{Mittiga:2018}.
    \item The stark variation in the effective field values $\Pi_\perp$ and $\Pi_\parallel$ experienced by different NVs located near the grain boundary suggest that there is a strong gradient of the strain field in this region. Strain gradients in a PCD sample was first shown by Trusheim \etal in \cite{Trusheim:2016} using wide-field microscopy technique. 
    \item We also performed high-resolution spectroscopy on single NVs located far away from the grain boundary ($>20 \mu m$). In this region, we found that there is a negligible  spatial variation of the effective field parameters (data not shown).
    \item We also performed continuous wave (CW) ODMR measurements on NV centers located on either side of the grain boundary. The CW ODMR spectra at zero magnetic field of single NVs present on one side were found to be positively shifted from 2.87 GHz central frequency, whereas the spectra for the NVs on the other side were found to be negatively shifted. One such region of the PCD sample is shown in figure \ref{fig:Positive_&_negative_shift_CW_ODMR_and_Scan}(b) where we investigated a set of single NV defects on either side of the grain boundary. {Figure} \ref{fig:Positive_&_negative_shift_CW_ODMR_and_Scan}(a) shows the CW ODMR spectrum of two NVs with same orientation on different sides of the grain boundary, showing positive and negative shift of the overall spectrum.
\end{itemize}
\section{Conclusion}
We have presented a study of the ground-state hyperfine spin level structure of NV defects in diamond in the presence of intrinsic effective fields and external magnetic fields. Apart from resolving hyperfine splitting using pulsed ESR spectroscopy, we have experimentally verified the previously unreported ESR features of mixing and splitting of the outer hyperfine transitions resulting from the interactions with the strain component of the effective field. We have observed good agreement between magnetic field-dependent ODMR fluorescence measurements and a simple theoretical model. These studies also enabled us to experimentally measure previously unobserved transition imbalances for the $m_I=\pm1$ hyperfine projections of the NV center. Finally, we have also developed a theoretical model that explores the interplay between the elliptical polarization of the microwave drive and the polarization response of the hyperfine spin transitions. Thus, we have investigated an unexplored regime $\Pi_{\perp}\geq A_{HF}$ in which the effects due to the intrinsic effective field and the hyperfine interaction are of comparable order.\vspace{0.5 cm}\\
Our studies open the door to a number of interesting research directions. First, our studies provide an in-depth understanding of the polarization selectivity of the hyperfine resonances depending on the ellipticity of the microwave excitation. This is of particular relevance for polarization-selective microwave excitation of NV centers in diamond samples for zero-field sensing applications. Second, single NV ODMR spectra of untreated type-Ib diamond we report on here provide clear signatures of the split-peak imbalance of the central transition with improved signal-to-noise ratio. Laser excitation wavelength dependent ODMR studies of single NV centers in these samples could provide valuable insight into the properties of the NV$^-$-N$^+$ pairs in diamond \cite{Manson:2018}. Third, our studies demonstrate the viability of hyperfine resolved microwave spectroscopy of a single NV center for atomic-scale resolution strain imaging in diamond with high sensitivity, which represents an advance over the earlier work on wide-field strain imaging \cite{Trusheim:2016}. 
\ack
P.P. gratefully acknowledges financial support from SERB ECRA grant number ECR/2018/002276 and STARS MoE grant number MoE-STARS/STARS1/662. S.K. acknowledges the CSIR funding agency award number 09/1020(0202)/2020-EMR-I for providing SRF fellowship. We also acknowledge Qudi Software Suite \cite{Binder:2017} for experiment control. 
\section*{Conflict of interest declaration}
All the authors declare no conflict of interest.
\begin{appendix}
\appendix
\section{Arbitrarily (elliptically) polarized microwave fields}
\label{section:Arbitrarily_Polarized_Microwaves}
\begin{figure}[H]
    \centering
    \includegraphics[scale=0.5]{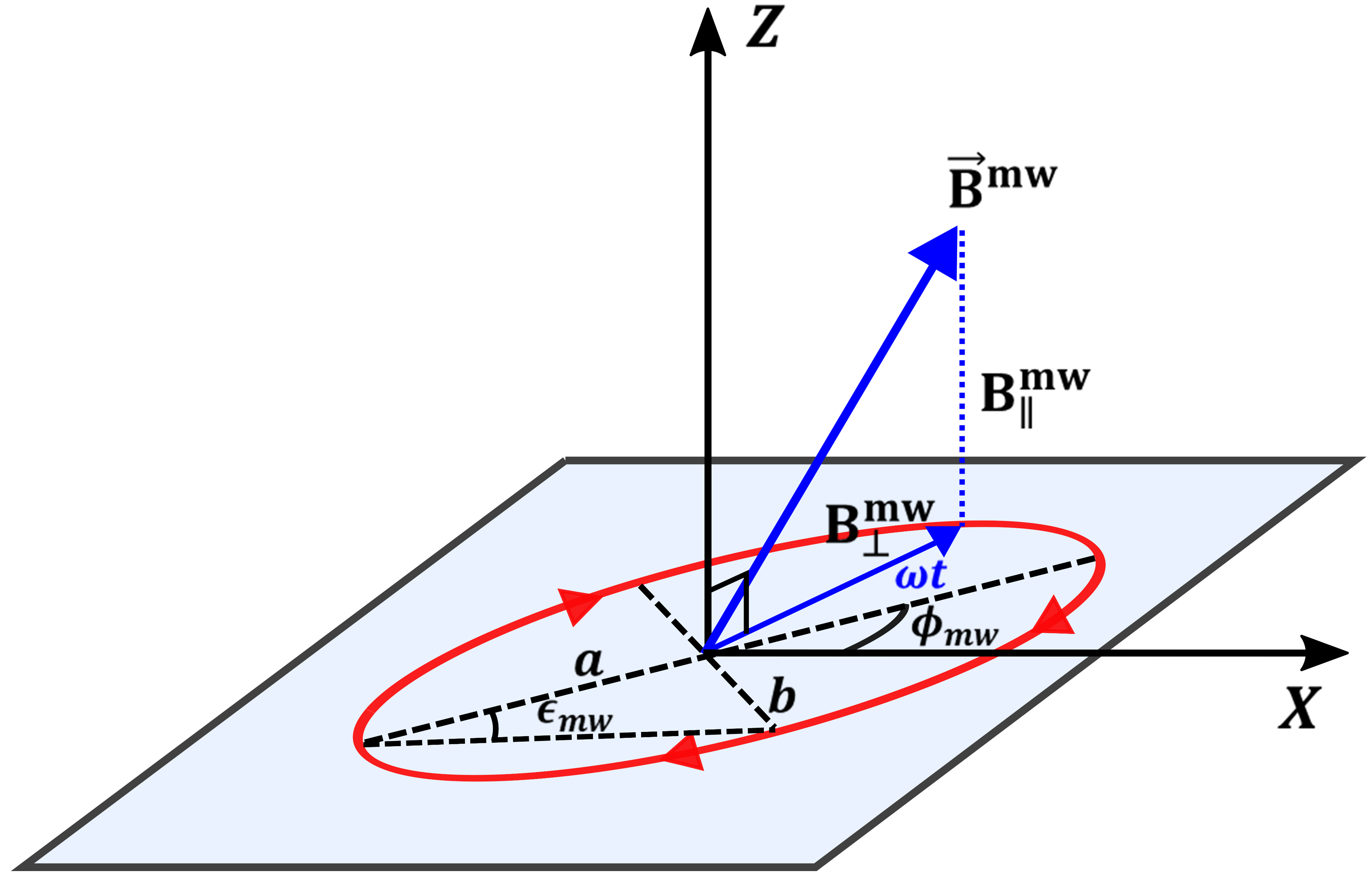}
    \caption{The elliptical path (red) traced by the MW field vector for any arbitrary (elliptical) polarization. $a$ and $b$ are the lengths of the semi-minor and semi-major axes of the ellipse, respectively.}
   \label{fig:Ellipse}
\end{figure}

\begin{figure}[htbp]
    \centering
    \includegraphics[scale=0.57]{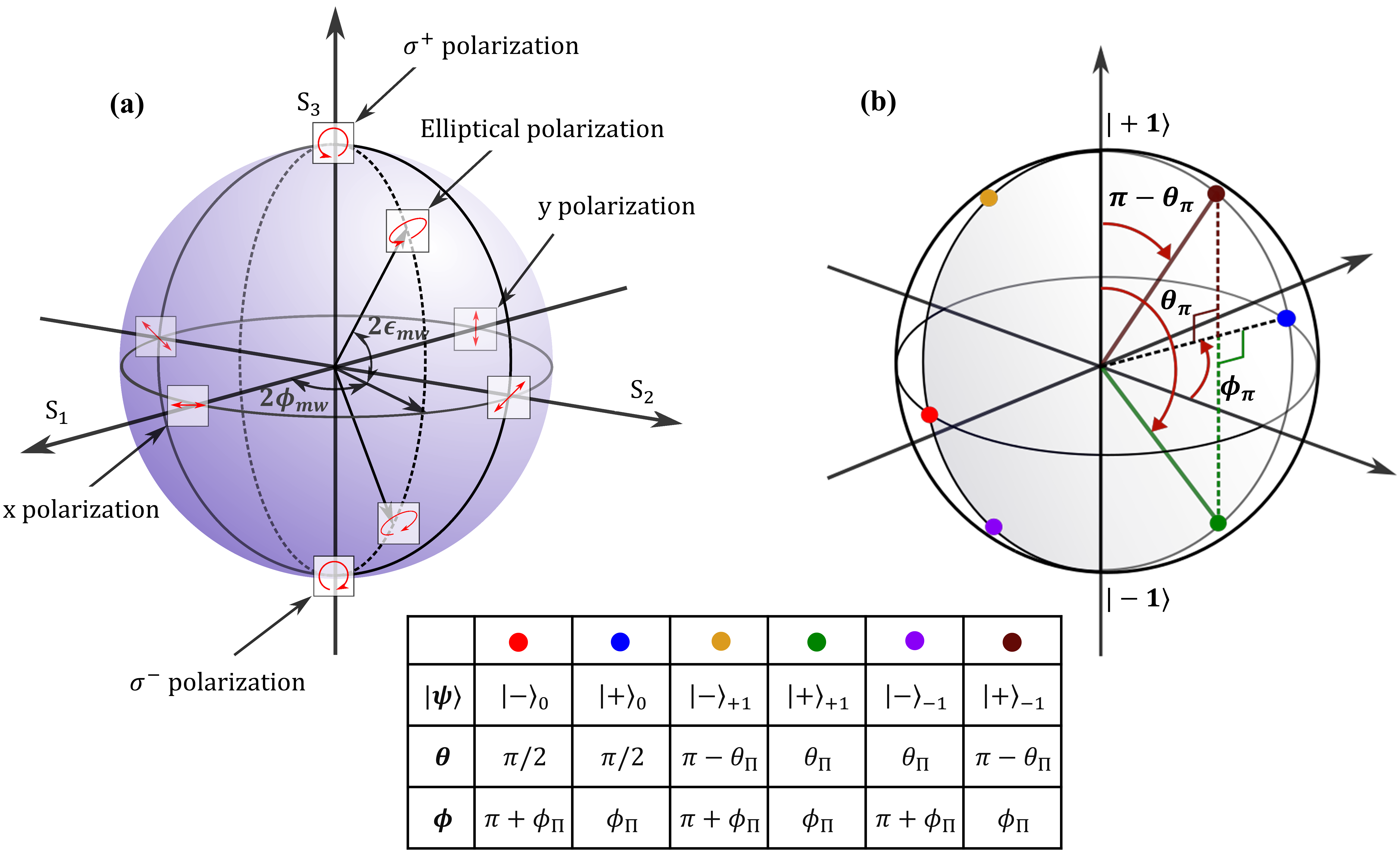}
    \caption{\textbf{(a)} The Poincare sphere representing the polarization states of the microwaves propagating in the $z$-direction with its magnetic field vector lying in the $x-y$ plane. The microwave polarization states are mapped to the Stokes parameters ($S_1,S_2,S_3$) on the surface of a unit Poincare sphere using the azimuthal angle $\phi_{mw}$ and the ellipticity angle $\epsilon_{mw}$. The antipodal points of the sphere represent a pair of mutually orthogonal polarization states. Points on the equatorial circle represent linearly polarized microwaves, and the points at the poles correspond to circularly polarized microwaves. All other points of the unit sphere indicate elliptical polarization states. \textbf{(b)} The Bloch sphere representation of the six NV electron spin states described in equations (\ref{eqn:All_main_eigenstates}) of the main text. The Bloch sphere is spanned by the two states $\ket{\pm1}$. Any general superposition state of $\ket{+1}$ and $\ket{-1}$ can be defined by the equation $\ket{\psi(\theta,\phi)} = \cos{\left(\frac{\theta}{2}\right)}\ket{+1}+e^{i\phi}\sin{\left(\frac{\theta}{2}\right)}\ket{-1}$, where $\theta$ and $\phi$ are the polar and azimuthal angles respectively. The transverse effective field experienced by a single NV can be parametrized using the angles $\theta_{\Pi}$ and $\phi_{\Pi}$. The six colored dots, corresponding to the six possible electron spin states, are present on the same great circle of the Bloch sphere. The transitions corresponding to the two states lying on the equator exhibit linearly polarized response, whereas the response is elliptically polarized for the remaining four states. }
    \label{fig:Poincare}
\end{figure}

In order to achieve full microwave polarization control of the coupling strength between the $\ket{0}_{m_I}$ and $\ket{\pm}_{m_I}$ states, we theoretically investigate the magnetic resonance spectrum obtained by driving the electron spin transitions from the $\ket{0}_{m_I}$ to the $\ket{\pm}_{m_I}$ states using an arbitrarily (elliptically) polarized microwave field $\vb*{B}^{mw}(t)$ (see figure \ref{fig:Poincare}). Here, we note that the $z$ component of the microwave field $B_{\parallel}^{mw}$ does not induce magnetic dipole transitions between the states $\ket{0}_{m_I}$ and $\ket{\pm}_{m_I}$. The most general expression for the microwave field vector $\vb*{B}^{mw}_{\perp}(t)$ perpendicular to the NV symmetry axis can be written using the parameters $\phi_{mw}$ and $\epsilon_{mw}$ as \cite{book:Auzinsh:2010}
\begin{align}
\label{Appendix:general_MW_field1}
\vb*{B}^{mw}_{\perp}(t)&=B^{mw}_{\perp}[\cos{\epsilon_{mw}}(\cos{\phi_{mw}}\vb*{e}_x+\sin{\phi_{mw}}\vb*{e}_y)\cos{\omega t} \nonumber\\
&+\sin{\epsilon_{mw}}(-\sin{\phi_{mw}}\vb*{e}_x+\cos{\phi_{mw}}\vb*{e}_y)\sin{\omega t}]
\end{align}
where the parameter $\epsilon_{mw}=\pm\arctan{\left(\frac{b}{a}\right)}$ is known as the ellipticity angle as shown in \cref{fig:Ellipse}. The time evolution of the magnetic field vector traces out an ellipse in the $xy$ plane with $a$ and $b$ being the length of the semi-major and semi-minor axes of the ellipse, and the parameter $\phi_{mw}$ is the angle the major axis of the ellipse makes with the $x$-axis. By introducing the parameter $\lambda_{mw}$, we can also parametrize the microwave magnetic field $\vb*{B}^{mw}_{\perp}(t)$ in terms of the ratio of the amplitudes of the left-circularly polarized microwaves ($\sigma^{+}$) and the right-circularly polarized microwaves ($\sigma^{-}$) as
\begin{align}
\label{Appendix:general_MW_field2}
    \vb*{B}^{mw}_{\perp}(t) &=  \vb*{B}_{\perp,\sigma^{+}}^{mw}(t)+\vb*{B}_{\perp,\sigma^{-}}^{mw}(t) \nonumber\\
    &=\cos(\frac{\pi}{4}-\epsilon_{mw})  \bigg(\frac{B^{mw}_{\perp}}{\sqrt{2}}[\cos(\omega t+\phi_{mw})\vb*{e}_x+\sin(\omega t+\phi_{mw})\vb*{e}_y]\\
    &  +\lambda_{mw}\frac{B^{mw}_{\perp}}{\sqrt{2}}[\cos(\omega t-\phi_{mw})\vb*{e}_x-\sin(\omega t-\phi_{mw})\vb*{e}_y] \bigg)\nonumber
\end{align}
where the parameter $\lambda_{mw}$ is related to $\epsilon_{mw}$ by $\lambda_{mw}=\tan{(\frac{\pi}{4}}-\epsilon_{mw})$.

Three special cases are of particular significance according to the equations (\ref{Appendix:general_MW_field1}) and (\ref{Appendix:general_MW_field2}): \\
(a) Linearly polarized microwaves corresponding to $\epsilon_{mw}=0$ or $\lambda_{mw}=1$ gives: 
\begin{subequations}
\begin{align}
\label{eqn:linearly_polarized_B}
\vb*{B}^{mw}_{\perp}(t) &=B^{mw}_{\perp}(\cos{\phi_{mw}}\vb*{e}_x+\sin{\phi_{mw}}\vb*{e}_y)\cos{\omega t},
\end{align}
(b) Left-circularly polarized microwaves ($\sigma^{+}$) corresponding to $\lambda_{mw}=0$ or $\epsilon_{mw}=\frac{\pi}{4}$ gives:  
\begin{align}
\label{eqn:Left_circularly_polarized_B}
    \vb*{B}^{mw}_{\perp}(t) &=  \frac{B^{mw}_{\perp}}{\sqrt{2}}[\cos(\omega t+\phi_{mw})\vb*{e}_x+\sin(\omega t+\phi_{mw})\vb*{e}_y],
\end{align}
(c) Right-circularly polarized microwaves ($\sigma^{-}$) corresponding to $\epsilon_{mw} = -\frac{\pi}{4}$ give: 
\begin{align}
\label{eqn:Right_circularly_polarized_B}
    \vb*{B}^{mw}_{\perp}(t) &=  \frac{B^{mw}_{\perp}}{\sqrt{2}}[\cos(\omega t-\phi_{mw})\vb*{e}_x-\sin(\omega t-\phi_{mw})\vb*{e}_y].
\end{align}
\end{subequations}
\section{Derivation of the rotating frame Hamiltonian }
The general Hamiltonian of the magnetic dipole moment $\vb*{\hat{\mu}}=-\gamma\vb*{\hat{S}}$ interacting with the elliptically polarized microwave control field is described as follows:
\begin{align}
    \mathcal{\hat{H}}^{mw}(t) &= \gamma(\vb*{B}_{\perp,\sigma^{+}}^{mw}(t)+\vb*{B}_{\perp,\sigma^{-}}^{mw}(t))\cdot\vb*{\hat{S}}\nonumber\\
    &= \cos{\left(\frac{\pi}{4}-\epsilon_{mw}\right)}\bigg(\frac{\Omega}{\sqrt{2}}[\cos(\omega t+\phi_{mw})\hat{S}_x+\sin(\omega t+\phi_{mw})\hat{S}_y]\nonumber\\
     &+\lambda_{mw}\;\frac{\Omega}{\sqrt{2}}[\cos(\omega t-\phi_{mw})\hat{S}_x-\sin(\omega t-\phi_{mw})\hat{S}_y]\bigg)
\end{align}
where $\Omega=\gamma B^{mw}_{\perp}$ is the Rabi frequency and $\omega$ is the frequency of the microwave. In the $\hat{S}_z$ basis $\{\ket{+1},\ket{0},\ket{-1}\}$, $\hat{S}_x$ and $\hat{S}_y$ being the $x$ and $y$ components of the spin-1 vector operator $\vb*{\hat{S}}$ are described as 
\begin{align}
    \hat{S}_x = \frac{1}{\sqrt{2}} 
    \begin{bmatrix}
    0 & 1 & 0\\
    1 & 0 & 1\\
    0 & 1 & 0
    \end{bmatrix},\,\, 
    \hat{S}_y = \frac{1}{\sqrt{2}}\begin{bmatrix}
    0 & -i & 0\\
    i & 0 & -i\\
    0 & i & 0
    \end{bmatrix}
\end{align}
The matrix representation for the Hamiltonian in the basis $\{\ket{+1},\ket{0},\ket{-1}\}$ is given by
\begin{align}
     \mathcal{\hat{H}}^{mw}(t) &=\cos{\left(\frac{\pi}{4}-\epsilon_{mw}\right)} \frac{\Omega}{{2}}
      \begin{bmatrix}
    0& e^{-i(\omega t+\phi_{mw} )} & 0\\
    e^{i(\omega t+\phi_{mw} )}& 0 & e^{-i(\omega t+\phi_{mw} )}\\
    0&  e^{i(\omega t+\phi_{mw} )} & 0
    \end{bmatrix} \nonumber\\ \nonumber \\
   & +\cos{\left(\frac{\pi}{4}-\epsilon_{mw}\right)} \frac{\Omega}{{2}} 
   \begin{bmatrix}
    0& \lambda_{mw}\;e^{i(\omega t-\phi_{mw} )}& 0\\
    \lambda_{mw}\;e^{-i(\omega t-\phi_{mw} )}& 0 & \lambda_{mw}\;e^{i(\omega t-\phi_{mw} )}\\
    0&  \lambda_{mw}\;e^{-i(\omega t-\phi_{mw} )} & 0
    \end{bmatrix}
\end{align}
By making a transformation to the rotating frame $\hat{U}=e^{-i\omega t\hat{S}_z^2}$, the transformed Hamiltonian has the form
\begin{align}
    \mathcal{\hat{H}}^{'mw}(t) &= \hat{U}^{\dagger} \mathcal{\hat{H}}^{mw}(t) \hat{U} \nonumber\\
    &= \cos{\left(\frac{\pi}{4}-\epsilon_{mw}\right)} \frac{\Omega}{{2}}\begin{bmatrix}
    0& e^{-i\phi_{mw}} & 0\\
    e^{i\phi_{mw}}& 0 & e^{-i(2\omega t+\phi_{mw})}\\
    0 & e^{i(2\omega t+\phi_{mw})} & 0
    \end{bmatrix} \nonumber\\ 
    &+\sin{\left(\frac{\pi}{4}-\epsilon_{mw}\right)} \frac{\Omega}{{2}} 
    \begin{bmatrix}
    0& e^{-i(\phi_{mw}-2\omega t)} & 0\\
    e^{i(\phi_{mw}-2\omega t)}& 0 & e^{-i\phi_{mw}}\\
    0 & e^{i\phi_{mw}}& 0
    \end{bmatrix}
\end{align}
where we used the relation $\lambda_{mw}=\tan{(\frac{\pi}{4}}-\epsilon_{mw})$. Neglecting all the time dependent terms rotating at $2\omega$ (rotating wave approximation), the final MW Hamiltonian \cite{Fang:2013} takes the form 
\begin{align}
     \mathcal{\hat{H}}^{'mw} &=  \frac{\Omega}{{2}}\begin{bmatrix}
     0 & \cos{\left(\frac{\pi}{4}-\epsilon_{mw}\right)}e^{-i\phi_{mw}}& 0 \\
     \cos{\left(\frac{\pi}{4}-\epsilon_{mw}\right)}e^{i\phi_{mw}}& 0 & \sin{\left(\frac{\pi}{4}-\epsilon_{mw}\right)}e^{-i\phi_{mw}}\\
     0& \sin{\left(\frac{\pi}{4}-\epsilon_{mw}\right)}e^{i\phi_{mw}}& 0
     \end{bmatrix} \nonumber\\\nonumber\\
&= \frac{\Omega}{{2}}\left[\cos{\left(\frac{\pi}{4}-\epsilon_{mw}\right)}e^{-i\phi_{mw}}|1\rangle\bra{0}+\sin{\left(\frac{\pi}{4}-\epsilon_{mw}\right)}e^{i\phi_{mw}}|-1\rangle\bra{0}\right] \nonumber\\
&+ \frac{\Omega}{{2}}\left[\sin{\left(\frac{\pi}{4}-\epsilon_{mw}\right)}e^{-i\phi_{mw}}\ket{0}\langle-1|+\cos{\left(\frac{\pi}{4}-\epsilon_{mw}\right)}e^{i\phi_{mw}}\ket{0}\langle1|\right]  \nonumber\\\nonumber\\
  &=\frac{\Omega}{2}\left[\cos{\left(\frac{\pi}{4}-\epsilon_{mw}\right)}\;e^{-i \phi_{mw}}|1\rangle+\sin{\left(\frac{\pi}{4}-\epsilon_{mw}\right)}\;e^{i\phi_{mw}}|-1\rangle\right] \bra{0} + \text{H.c.}
\end{align}
where H.c. is the Hermitian conjugate. 
\section{Magnetic dipole transition strengths}
\label{subsection:Magnetic_Dipole_Transition_Strengths}
At zero field, the Rabi frequencies corresponding to the magnetic dipole transitions $\ket{0}_{m_I}\rightarrow \ket{\pm}_{m_I}$ are
\begin{subequations}
\begin{equation}
     \Omega_{\pm,m_I}=2\pi\abs{\bra{\pm,m_I}\mathcal{\hat{H}}^{'mw}\ket{0,m_I}}.\nonumber
\end{equation}
So, the six hyperfine spin transition strengths under driving by an arbitrarily polarized microwave field are given by
\label{eqn:All_transition_strengths}
\begin{align}
    \mathcal{A}_{-,0}&= \Omega_{-,0}^2= (2 \pi)^2 \left(\frac{\Omega}{2}\right)^2 \left| \frac{1}{\sqrt{2}}\cos{\left(\frac{\pi}{4}-\epsilon_{mw}\right)}- \frac{1}{\sqrt{2}}\sin{\left(\frac{\pi}{4}-\epsilon_{mw}\right)}\,e^{i(2\phi_{mw}-\phi_\Pi)}\right|^2 \nonumber\\
      &=(2 \pi)^2 \left(\frac{\Omega}{2}\right)^2W_{-,0} \label{eqn:Transition_strength_minus_mi0-},\\
    \mathcal{A}_{+,0}&= \Omega_{+,0}^2= (2 \pi)^2 \left(\frac{\Omega}{2}\right)^2 \left| \frac{1}{\sqrt{2}}\cos{\left(\frac{\pi}{4}-\epsilon_{mw}\right)}+ \frac{1}{\sqrt{2}}\sin{\left(\frac{\pi}{4}-\epsilon_{mw}\right)}\,e^{i(2\phi_{mw}-\phi_\Pi)}\right|^2\nonumber\\
      &=(2 \pi)^2 \left(\frac{\Omega}{2}\right)^2W_{+,0} \label{eqn:Transition_strength_plus_mi0},\\ \nonumber\\
     \mathcal{A}_{-,+1}&= \Omega_{-,+1}^2= (2 \pi)^2 \left(\frac{\Omega}{2}\right)^2 \left|\sin{\left(\frac{\theta_{\Pi}}{2}\right)} \cos{\left(\frac{\pi}{4}-\epsilon_{mw}\right)}- \cos{\left(\frac{\theta_{\Pi}}{2}\right)} \sin{\left(\frac{\pi}{4}-\epsilon_{mw}\right)}\,e^{i(2\phi_{mw}-\phi_\Pi)}\right|^2 \label{eqn:Transition_strength_minus_mi+1} \nonumber\\
      &=\left(2 \pi\right)^2 \left(\frac{\Omega}{2}\right)^2W_{-,+1},\\
    \mathcal{A}_{+,+1}&= \Omega_{+,+1}^2= (2 \pi)^2 \left(\frac{\Omega}{2}\right)^2 \left|\cos{\left(\frac{\theta_{\Pi}}{2}\right)} \cos{\left(\frac{\pi}{4}-\epsilon_{mw}\right)}+ \sin{\left(\frac{\theta_{\Pi}}{2}\right)} \sin{\left(\frac{\pi}{4}-\epsilon_{mw}\right)}\,e^{i(2\phi_{mw}-\phi_\Pi)}\right|^2 \label{eqn:Transition_strength_plus_mi+1}\nonumber\\  
      &=(2 \pi)^2 \left(\frac{\Omega}{2}\right)^2W_{+,+1},\\ \nonumber\\
     \mathcal{A}_{-,-1}&= \Omega_{-,-1}^2= (2 \pi)^2 \left(\frac{\Omega}{2}\right)^2 \left|\cos{\left(\frac{\theta_{\Pi}}{2}\right)} \cos{\left(\frac{\pi}{4}-\epsilon_{mw}\right)}- \sin{\left(\frac{\theta_{\Pi}}{2}\right)} \sin{\left(\frac{\pi}{4}-\epsilon_{mw}\right)}\,e^{i(2\phi_{mw}-\phi_\Pi)}\right|^2 \label{eqn:Transition_strength_minus_mi-1}\nonumber\\
      &=\left(2 \pi\right)^2 \left(\frac{\Omega}{2}\right)^2W_{-,-1},\\
     \mathcal{A}_{+,-1}&= \Omega_{+,-1}^2= (2 \pi)^2 \left(\frac{\Omega}{2}\right)^2 \left|\sin{\left(\frac{\theta_{\Pi}}{2}\right)} \cos{\left(\frac{\pi}{4}-\epsilon_{mw}\right)}+ \cos{\left(\frac{\theta_{\Pi}}{2}\right)} \sin{\left(\frac{\pi}{4}-\epsilon_{mw}\right)}\,e^{i(2\phi_{mw}-\phi_\Pi)}\right|^2\nonumber\\
      &=(2 \pi)^2 \left(\frac{\Omega}{2}\right)^2W_{+,-1}, \label{eqn:Transition_strength_plus_mi-1}
\end{align}
\end{subequations}
where
\begin{align}
    \cos{\theta_{\Pi}}&= \frac{-|A_{HF}|}{\sqrt{\Pi_{\perp}^2+\left(A_{HF}\right)^2}}\nonumber
\end{align}
and $W_{\pm,m_I}$ are the normalized transition strengths. Therefore, the transition imbalance of the inner $m_I=0$ transitions is given by
\begin{subequations}
\begin{align}
    \mathcal{I}_{inner}&=\frac{\mathcal{A}_{0,+}- \mathcal{A}_{0,-}}{\mathcal{A}_{0,+}+ \mathcal{A}_{0,-}} \nonumber\\
    &= \sin{\left[2\left(\frac{\pi}{4}-\epsilon_{mw}\right)\right]} \cos{(2\phi_{mw}-\phi_\Pi)}, \label{eqn:Inner_Imbalance_for_elliptical_polarisation}
\end{align}
and the transition imbalance of the outer $m_I=\pm1$ transitions is given by
\begin{align}
\mathcal{I}_{outer}&=\frac{\mathcal{A}_{+,\pm 1}- \mathcal{A}_{-,\pm 1}}{\mathcal{A}_{+,\pm 1}+ \mathcal{A}_{-,\pm 1}}\nonumber\\
    &= \pm\cos{\theta_{\Pi}} \cos{\left[2\left(\frac{\pi}{4}-\epsilon_{mw}\right)\right]}+ \sin{\theta_{\Pi}} \sin{\left[2\left(\frac{\pi}{4}-\epsilon_{mw}\right)\right]} \cos{(2\phi_{mw}-\phi_\Pi)}. \label{eqn:Outer_Imbalance_for_elliptical_polarisation}
\end{align}
\end{subequations}
\begin{table}[h]
\renewcommand{\arraystretch}{1.8}
    \caption{Hyperfine transition strengths at zero magnetic field as a function of the polarization of the microwaves.}
    \label{tab:Transition_strength_table}
\begin{center}
\begin{tabular} { | M{8em}| M{5cm}| M{3cm} | M{3cm} | } 
 \hline
    & Linear polarization & $\sigma^{+}$ polarization &  $\sigma^{-}$ polarization\\ 
  \hline
  
 $\ket{0,0} \rightarrow \ket{-,0}$ & $\frac{1}{2}(1-\cos (2\phi_{mw}-\phi_{\Pi}))$ & 0.500 & 0.500\\ 
  \hline
  $\ket{0,0} \rightarrow \ket{+,0}$ & $\frac{1}{2}(1+\cos (2\phi_{mw}-\phi_{\Pi}))$ & 0.500 & 0.500  \\ 
  \hline
   $\ket{0,+1} \rightarrow \ket{-,+1}$ &$\frac{1}{2}(1-\sin{\theta_{\Pi}}\,\cos (2\phi_{mw}-\phi_{\Pi}))$ &$\sin^2{\frac{\theta_{\Pi}}{2}}$ & $\cos^2{\frac{\theta_{\Pi}}{2}}$\\
  \hline
  $\ket{0,+1} \rightarrow \ket{+,+1}$ & $\frac{1}{2}(1+\sin{\theta_{\Pi}}\,\cos (2\phi_{mw}-\phi_{\Pi}))$ & $\cos^2{\frac{\theta_{\Pi}}{2}}$  & $\sin^{2}{\frac{\theta_{\Pi}}{2}}$ \\
  \hline
   $\ket{0,-1} \rightarrow \ket{-,-1}$ & $\frac{1}{2}(1-\sin{\theta_{\Pi}}\,\cos (2\phi_{mw}-\phi_{\Pi}))$ & $\cos^2{\frac{\theta_{\Pi}}{2}}$  & $\sin^{2}{\frac{\theta_{\Pi}}{2}}$ \\
  \hline
   $\ket{0,-1} \rightarrow \ket{+,-1}$ & $\frac{1}{2}(1+\sin{\theta_{\Pi}}\,\cos (2\phi_{mw}-\phi_{\Pi}))$ & $\sin^2{\frac{\theta_{\Pi}}{2}}$  & $\cos^2{\frac{\theta_{\Pi}}{2}}$ \\
  \hline
\end{tabular}
\end{center}
\end{table}
\begin{table}[htbp]
\renewcommand{\arraystretch}{1.8}
    \caption{Hyperfine transition strengths at zero magnetic field for two different ellipticity values of the microwave polarization. For these two ellipticity values, the individual $m_I=\pm1$ outer transitions can be selectively excited by tuning the azimuthal angle of the elliptically polarized MW drive relative to the azimuthal angle of the effective field. }
\begin{center}
\begin{tabular} { | M{8em}| M{6cm} | M{5.5cm} | } 
 \hline
    & $\epsilon = \frac{\pi}{4}-\frac{\theta_{\Pi}}{2}$ & $\epsilon =-(\frac{\pi}{4}-\frac{\theta_{\Pi}}{2})$\\ 
  \hline
  $\ket{0,0} \rightarrow \ket{-,0}$ & $\frac{1}{2}(1-\sin{\theta_\Pi}\cos (2\phi_{mw}-\phi_{\Pi}))$&  $\frac{1}{2}(1-\sin{\theta_\Pi}\cos (2\phi_{mw}-\phi_{\Pi}))$\\ 
  \hline
  $\ket{0,0} \rightarrow \ket{+,0}$ & $\frac{1}{2}(1+\sin{\theta_\Pi}\cos (2\phi_{mw}-\phi_{\Pi}))$ & $\frac{1}{2}(1+\sin{\theta_\Pi}\cos (2\phi_{mw}-\phi_{\Pi}))$ \\ 
  \hline
   $\ket{0,+1} \rightarrow \ket{-,+1}$ &$\frac{1}{2}\sin^2{\theta_\Pi}(1-\cos (2\phi_{mw}-\phi_{\Pi}))$& $1-\frac{1}{2}\sin^2{\theta_\Pi}(1+\cos (2\phi_{mw}-\phi_{\Pi}))$ \\
  \hline
  $\ket{0,+1} \rightarrow \ket{+,+1}$ & $1-\frac{1}{2}\sin^2{\theta_\Pi}(1-\cos (2\phi_{mw}-\phi_{\Pi}))$& $\frac{1}{2}\sin^2{\theta_\Pi}(1+\cos (2\phi_{mw}-\phi_{\Pi}))$ \\
  \hline
   $\ket{0,-1} \rightarrow \ket{-,-1}$ & $1-\frac{1}{2}\sin^2{\theta_\Pi}(1+\cos (2\phi_{mw}-\phi_{\Pi}))$& $\frac{1}{2}\sin^2{\theta_\Pi}(1-\cos (2\phi_{mw}-\phi_{\Pi}))$ \\
  \hline
   $\ket{0,-1} \rightarrow \ket{+,-1}$ & $\frac{1}{2}\sin^2{\theta_\Pi}(1+\cos (2\phi_{mw}-\phi_{\Pi}))$&$1-\frac{1}{2}\sin^2{\theta_\Pi}(1-\cos (2\phi_{mw}-\phi_{\Pi}))$ \\
  \hline
\end{tabular}
\end{center}
\end{table}
\begin{figure}[h]
    \centering
    \includegraphics[scale=0.9]{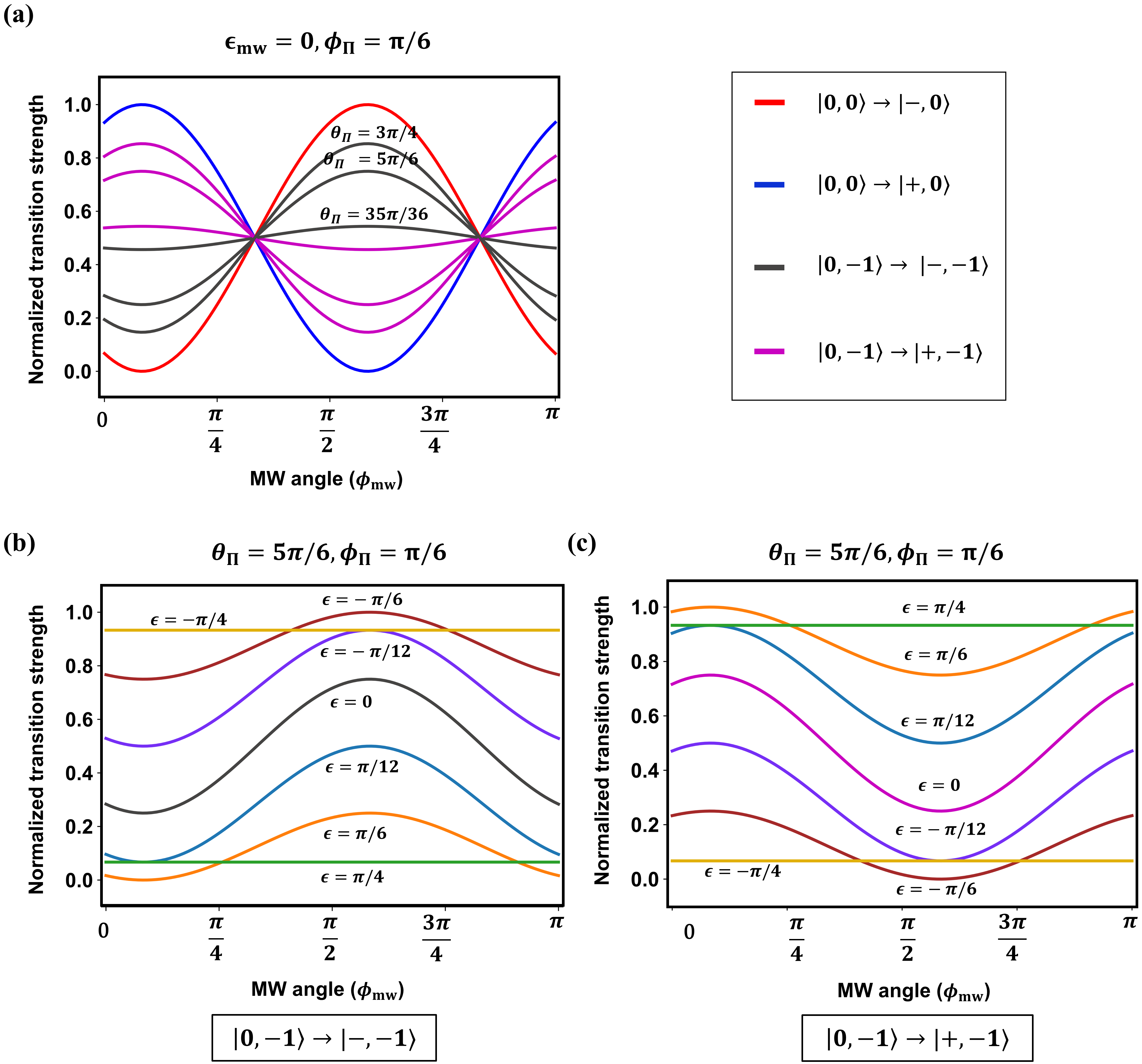}    \caption{\textbf{(a)} Plots of the normalized transition strengths $W_{-,m_I}$ and $W_{+,m_I}$ for linear microwave polarization, as a function of the MW angle $\phi_{mw}$ for various values of the angle $\theta_\Pi$ and given $\phi_\Pi = \frac{\pi}{6}$. \textbf{(b)} Normalized transition strength of outer transition $\ket{0,-1}\rightarrow\ket{-,-1}$ for elliptically polarized MW field as a function of MW angle for various $\epsilon$, where $\phi_\Pi = \frac{\pi}{6}$ and $\theta_\Pi = \frac{5\pi}{6}$. It is clear from the figure that the normalized transition strength may approach 1 and 0 depending on the ellipticity value. This suggests that a fully bright and fully dark state is achievable for outer states also when excited with an appropriately polarized MW field. \textbf{(c)} Normalized transition strength of outer transition $\ket{0,-1}\rightarrow\ket{+,-1}$.}
    \label{fig:Transition_strength_for_different_theta_and_epsilon}
\end{figure}

Since the transition strengths of the $m_I = +1$ and  $ m_I = -1$ substates sum together, the transition strengths observed for the outer transitions will be 
\begin{subequations}
\begin{align}
    \mathcal{A}_- & = \mathcal{A}_{-,+ 1} + \mathcal{A}_{-,- 1} \nonumber\\
    & = (2 \pi)^2 \left(\frac{\Omega}{2}\right)^2\left(1- \sin{\left[2\left(\frac{\pi}{4} -\epsilon\right)\right]} \sin{\theta_\Pi}\cos{(2\phi_{mw}-\phi_\Pi)}\right), 
\end{align}

and 
\begin{align}
    \mathcal{A}_+ & = \mathcal{A}_{+,- 1} + \mathcal{A}_{+,+ 1} \nonumber\\
    & = (2 \pi)^2 \left(\frac{\Omega}{2}\right)^2 \left(1+ \sin{\left[2\left(\frac{\pi}{4} -\epsilon\right)\right]} \sin{\theta_\Pi}\cos{(2\phi_{mw}-\phi_\Pi)}\right). 
\end{align}
So the imbalance observed for the outer states will be 
\begin{align}
    \mathcal{I}_{outer} &= (\mathcal{A}_+ -\mathcal{A}_-)/(\mathcal{A}_+ + \mathcal{A}_-)\nonumber\\
    & = \sin{\theta_\Pi}\sin{\left[2\left(\frac{\pi}{4}-\epsilon\right)\right] \cos{(2\phi_{mw}-\phi_\Pi)}}.
\end{align}
\end{subequations}
The relation between the Rabi frequencies at zero magnetic field can be obtained directly from the \cref{eqn:All_transition_strengths} and is given as: 
\begin{align}
    \Omega_1^2+\Omega_4^2 = \Omega_2^2+\Omega_3^2
\end{align}
where the subscripts 1, 2, 3 and 4 correspond to the transitions marked in the figure \ref{fig:polarisation_port1_and_2_and_polarization_response}(c) of the main text. 
\end{appendix}
\bibliography{References}

\begin{thebibliography}{10}
\bibitem{Footnote1}
 Please note that the pulsed odmr spectrum shown in figure
  \ref{fig:PCD+HPHT+3dips+5dips}(c) was obtained by conducting measurements on
  another nv defect experiencing a different effective field environment in the
  polycrystalline diamond sample under study.
\end{thebibliography}
\end{document}